\documentclass[aps,pre,preprint]{revtex4}
\usepackage{graphicx}
\usepackage{amsfonts}
\usepackage{amssymb}
\usepackage{amsmath}

\begin{document}

\title{Modulated Amplitude Waves in Collisionally Inhomogeneous
Bose-Einstein Condensates}
\author{Mason A. Porter}
\affiliation{Department of Physics and Center for the Physics of Information, California
Institute of Technology, Pasadena, CA 91125, USA}
\author{P. G. Kevrekidis}
\affiliation{Department of Mathematics and Statistics, University of Massachusetts,
Amherst MA 01003, USA}
\author{Boris A. Malomed}
\affiliation{Department of Interdisciplinary Studies, School of Electrical Engineering,
Faculty of Engineering, Tel Aviv University, Tel Aviv 69978, Israel}
\author{D. J. Frantzeskakis}
\affiliation{Department of Physics, University of Athens, Panepistimiopolis, Zografos,
Athens 15784, Greece}

\begin{abstract}
We investigate the dynamics of an effectively one-dimensional
Bose-Einstein condensate (BEC) with scattering length $a$
subjected to a spatially periodic modulation, $a=a(x)=a(x+L)$.
This ``collisionally inhomogeneous" BEC is described by a
Gross-Pitaevskii (GP) equation whose nonlinearity coefficient is a
periodic function of $x$. We transform this equation into a GP equation with constant coefficient $a$ and an additional effective potential and study a class of extended wave
solutions of the transformed equation.
For weak underlying inhomogeneity, the effective potential takes a
form resembling a superlattice, and the amplitude dynamics of the solutions of the constant-coefficient GP equation obey a nonlinear generalization of the
Ince equation. In the small-amplitude limit, we use averaging to
construct analytical solutions for modulated amplitude waves
(MAWs), whose stability we subsequently examine using both
numerical simulations of the original GP equation and fixed-point
computations with the MAWs as numerically exact solutions. We show
that ``on-site" solutions, whose maxima correspond to maxima of
$a(x)$, are significantly more stable than their ``off-site" counterparts.
\end{abstract}

\maketitle

PACS: 05.45.-a, 03.75.Lm, 05.30.Jp, 05.45.Ac


Keywords: Bose-Einstein condensates, periodic potentials, multiple-scale
perturbation theory, Hamiltonian systems

\section{Introduction}

Among the most fundamental models studied widely in applied mathematics and
employed for the description of an extremely large variety of physical
phenomena are nonlinear Schr\"{o}dinger (NLS)\ equations \cite{sulem}.
Applications of NLS equations have become even more prominent in recent years
due to enormous theoretical and experimental progress that has taken place
in studies of Bose-Einstein condensates (BECs) \cite {book2} and nonlinear optics \cite{book1}. Through the formal similarity between coherent matter waves and electromagnetic waves, research on these topics has become closely connected, and progress in one area frequently also benefits the other. The
cross-fertilization between these two fields is extremely important not only
from a theoretical perspective but also for applications. For example,
coherent matter waves can be manipulated using devices such as atom
chips \cite{Folman} whose design was initially suggested by previously-developed
optical counterparts.

In the mean-field approximation, and at sufficiently low temperatures, the
dynamics of matter waves are accurately modeled by a cubic NLS
equation incorporating an external potential. In this context, it is called
the Gross-Pitaevskii (GP) equation \cite{book2},
\begin{equation}
	i\hbar {\Psi }_{t}=\left[ -\frac{\hbar ^{2}}{2m}\nabla ^{2}+\tilde{g}|\Psi
|^{2}+{\mathcal{V}}(\vec{r})\right] \Psi \,,  \label{GPE}
\end{equation}
where $m$ is the atomic mass, $\Psi (\vec{r},t)$ is the macroscopic wave
function of the condensate, normalized to the number of atoms $N$ (so that $%
\int |\Psi |^{2}d\mathbf{r}=N$), ${\mathcal{V}}(\vec{r})$ is the external
potential, and the effective interaction constant is $\tilde{g}=(4\pi \hbar
^{2}a/m)[1+O(\zeta ^{2})]$, where $a$ is the \textit{s}-wave scattering
length, and $\zeta ^{2}\equiv |\Psi |^{2}|a|^{3}$ measures the density of
the atomic gas \cite{book2}.

The properties of BECs -- including their shape, collective nonlinear
excitations (such as solitons and vortices), and fluctuations above the
mean-field level -- are determined by the sign and magnitude of the
scattering length $a$. Accordingly, one of the key tools used in current
studies of BECs relies on adjusting $a$ (and hence the nonlinearity
coefficient $\tilde{g}$ defined above). A well-known way to achieve this
goal is to tune an external magnetic field in the vicinity of a Feshbach
resonance \cite{Weiner03,feshbachNa,feshbachRb}. Alternatively, one can use
a Feshbach resonance induced by an optical \cite{Theis04} or dc electric
\cite{electric} field. In low-dimensional settings, one can also tune the
effective nonlinearity by changing the BEC's transversal confinement \cite%
{olshani,Napoli}.

Adjusting the scattering length \textit{globally} (i.e., modifying $a$ in a
spatially uniform manner) has been crucial to many experimental
achievements, including the formation of molecular condensates \cite%
{molecule} and probing the so-called BEC-BCS crossover \cite{becbcs}.
Additionally, recent theoretical studies have predicted that a spatially
uniform but time-periodic modulation of the scattering length, with $a(t)$
periodically changing its sign, can be used to stabilize attractive
condensates in two \cite{FRM1} and three \cite{FRMr} dimensions and thus help to create robust matter-wave solitons \cite{FRM2}.   [In the three-dimensional case, the Feshbach resonance-based technique should be combined with a quasi-one-dimensional (quasi-1D) periodic potential, known as an optical
lattice (OL).]  Other nonlinear waves besides solitons have also been studied in this context.  For example, it was predicted that spatially periodic or
quasiperiodic 2D patterns resembling the classical Faraday ripples in
hydrodynamics can be excited in a BEC via a nonlinear parametric resonance induced by
the spatially-uniform and temporally-periodic modulation of $a$ (but, in general,
without changing the sign of $a$) \cite{Kestutis}.

More recently, the possibility of varying the scattering length \textit{%
locally} (i.e., spatially) has also been proposed \cite%
{fka,fermi,g1,v1}. Such spatial dependence of the scattering length, which
can be implemented utilizing a spatially inhomogeneous external magnetic
field in the vicinity of a Feshbach resonance \cite{fermi,g1}, renders the
collisional dynamics inhomogeneous across the BEC.  Condensates with a
spatially inhomogeneous nonlinearity have recently attracted considerable
attention, as they are relevant to many significant applications, including
adiabatic compression of matter waves \cite{fka,g1}, Bloch oscillations of
matter-wave solitons \cite{g1}, atom lasers \cite{Peter,v1,v2}, enhancement
of transmittivity of matter waves through barriers \cite{g2,fka2}, and the
dynamics of matter waves in the presence of periodic or random spatial
variations of the scattering length \cite{fka3}. Additionally, while these works
chiefly concentrated on quasi-1D condensates, studies in a
quasi-two-dimensional (quasi-2D) setting were also recently reported \cite%
{malspace} (see also Ref.~\cite{Gadi} for a qualitatively similar model in a
different physical context).

In the present work, we consider an effectively one-dimensional BEC whose scattering length is subjected to a periodic variation: $a(x)=a(x+L)$ for some
period $L$.  We consider the case of no zero crossings, so that $a(x)$ always has the same sign.
Such a \textit{nonlinear lattice} can be realized experimentally (see, e.g.,
the diagram in Ref.~\cite{v2}), and it offers various possibilities for the
study of matter-wave solitons (as discussed in Refs.~\cite%
{fka3,malspace,fka4}). Here we focus on the dynamics of spatially extended
states (rather than solitons), which have not yet been considered in earlier
works in the context of nonlinear lattices. Extended states can be created
experimentally, as has been done, for example, in the setting of a BEC in an
optical superlattice \cite{quasibec}.

The analysis of the problem under consideration proceeds as follows. First,
we apply a transformation to the quasi-1D GP equation with the nonlinearity
coefficient periodically modulated in space in order to derive an effective
GP equation with $a=\mathrm{const}$. With this transformation, the original
inhomogeneity of $a(x)$ is mapped into an effective linear potential taking the form of a modified superlattice. We consider spatially extended solutions of the transformed GP equation in the form of \textit{modulated amplitude waves} (MAWs), whose slow-amplitude dynamics
we derive using an averaging method.  These analytical considerations, presented in Section II, are followed in Section III by an investigation of the dynamics and stability
of the MAWs using both direct numerical simulations of the
original GP equation and fixed-point computations with the MAWs as numerically exact solutions. We thereby show that ``on-site" solutions, whose maxima correspond to maxima of the
nonlinear lattice, tend to be more stable than ``off-site" solutions. We summarize our findings and present conclusions in Section IV.  We then discuss some technical details of the averaging procedure in an appendix (Section V).

\section{The Perturbed Gross-Pitaevskii Equation}

The 3D GP equation (\ref{GPE}) can be reduced to an effectively 1D form provided the transverse dimensions of the BEC are on the order of its
healing length and its longitudinal dimension is much larger than the
transverse ones \cite{gp1d,Napoli}. In this regime, the 1D approximation is
derived by averaging Eq.~(\ref{GPE}) in the transverse plane. One can
readily show (see, e.g., Ref.~\cite{g1}) that the resulting 1D GP equation
takes the dimensionless form
\begin{equation}
	iu_{t}=-\frac{1}{2}u_{xx}+g(x)|u|^{2}u+V(x)u\,,  \label{nls3}
\end{equation}
where we recall that nonlinearity coefficient $g(x)$ varies in space. In
Eq.~(\ref{nls3}), $u$ is the mean-field wave function (with density $|u|^{2}$
measured in units of the peak 1D density $n_{0}$), $x$ and $t$ are
normalized, respectively, to the healing length $\xi =\hbar /\sqrt{%
n_{0}|g_{1}|m}$ and $\xi /c$ (where $c=\sqrt{n_{0}|g_{1}|/m}$ is the
Bogoliubov speed of sound), and energy is measured in units of the
chemical potential $\mu =g_{1}n_{0}$. In the above expressions, $%
g_{1}=2\hbar \omega _{\perp }a_{0}$, where $\omega _{\perp }$ denotes the
confining frequency in the transverse direction, and $a_{0}$ is a
characteristic (constant) value of the scattering length relatively close to the
Feshbach resonance. Finally, $V(x)$ is the rescaled external trapping
potential, and the $x$-dependent nonlinearity is given by $g(x)=a(x)/a_{0}$,
where $a(x)$ is the spatially varying scattering length.

We will examine periodic modulations of the scattering length by assuming
\begin{equation}
	g(x)=g_{0}+V_{0}\sin ^{2}(\kappa x)\,,  \label{inhom}
\end{equation}%
where $V_{0}$ and $\kappa $ are, respectively, the amplitude and wavenumber
of the modulation. In experiments, this modulation mode can be induced by
a periodically patterned configuration of the external
(magnetic, optical, or electric) field that controls the Feshbach resonance. We
chiefly focus on the case of repulsive BECs, with $g_{0}>0$ and $V_{0}>0$.
For attractive BECs, for which $g_{0}<0$, one can also transform the
nonlinear partial differential equation (\ref{nls3}) into an effective GP equation with a
constant nonlinearity coefficient.

We now introduce the transformation, $v\equiv \sqrt{g(x)}u$, which casts
Eq.~(\ref{nls3}) in the form
\begin{eqnarray}
	iv_{t} &=&-\frac{1}{2}v_{xx}+|v|^{2}v+V(x)v+\hat{V}_{\mathrm{eff}}(x)v\,,
	\notag \\
	\hat{V}_{\mathrm{eff}}(x) &=&\frac{1}{2}\frac{f^{\prime \prime }}{f}-\frac{%
	\left( f^{^{\prime }}\right) ^{2}}{f^{2}}+\frac{f^{\prime }}{f}\frac{%
	\partial }{\partial x}\,,  \label{motion}
\end{eqnarray}%
where $f(x)\equiv \sqrt{g(x)}$, and $f^{\prime }\equiv df/dx$. Obviously,
this transformation applies only in the case when $g(x)$ does not cross zero.

If the scattering-length modulation is weak, i.e., $V_{0}\ll g_{0}$ in Eq.~(%
\ref{inhom}), then $\hat{V}_{\mathrm{eff}}$ can be approximated as a
superlattice potential (because it contains a second harmonic in addition to
the fundamental one) plus a first-derivative operator term:
\begin{equation}
	\hat{V}_{\mathrm{eff}}(x)=-\frac{3\kappa ^{2}V_{0}^{2}}{16g_{0}^{2}}+\frac{%
	\kappa ^{2}V_{0}}{2g_{0}}\cos (2\kappa x)+\frac{3\kappa ^{2}V_{0}^{2}}{%
	16g_{0}^{2}}\cos (4\kappa x)+\left[ \frac{V_{0}\kappa }{2g_{0}}\sin (2\kappa
	x)\right] \frac{\partial }{\partial x}\,.  
\label{effol}
\end{equation}
In this case, we define a small parameter, $\varepsilon \equiv \kappa
V_{0}/g_{0}$, evincing the fact that the $\cos (4\kappa x)$ harmonic in Eq.~(%
\ref{effol}) is a small correction to the fundamental one, $\cos (2\kappa x)$. Thus, to order $O(\epsilon )$, the effective potential (\ref{effol}) is a
modified lattice rather than a superlattice.

It is also worth mentionioning that in the case of a temporal modulation
of the scattering length, i.e., $g=g(t)$ in Eq. (\ref{nls3}), the same
transformation defined above can be used (i.e., $v\equiv \sqrt{g(t)}u$) provided $g(t)$ never vanishes. As is well known, the time-dependent coefficient in front of the nonlinear term is then translated into
a linear dissipative term on the right-hand side of the equation for $v(x,t)$
\cite{Manakov}.  This term has the time-dependent 
coefficient $-i(2g)^{-1}(dg/dt)v$. While this latter setting may be
of interest in its own right (perhaps especially when the dissipation
coefficient is constant so that the time dependence can be completely
factored out -- i.e., for $g(t)$ of an exponential form),
we do not pursue it further here. We remark that
if $g(t)$ is a periodic function, the transformed equation will
generate patterns with a periodically oscillating amplitude.  Finally, one can complicate the situation still further by considering simultaneous spatial and temporal modulations [i.e., $g = g(x,t)$], though such a configuration would be very difficult to controllably implement in experiments.

\section{Modulated Amplitude Waves}

We consider solutions to Eqs.~(\ref{motion},\ref{effol}) in the form of coherent structures given by the ansatz
\begin{equation}
	v(x,t)=R(x)\exp \left( i\left[ \theta (x)-\mu t\right] \right) \,.
\label{maw2}
\end{equation}
Such nonlinear waves, called \textit{modulated amplitude waves} (MAWs), have been
studied previously in BEC models with optical-lattice and superlattice
potentials \cite{map1,map2,map3,map4}.  Because of the spatial periodicity of the potential in Eq.~(\ref{motion}), MAWs in the linear limit yield the particular case of Bloch waves (of the transformed GP equation).  Generic Bloch wave functions are quasiperiodic in $x$; periodic ones lie at edges of Bloch
bands.  

Inserting Eq.~(\ref{maw2}) into Eq.~(\ref{motion}) and equating the real and
imaginary components of the resulting equation yields an ordinary
differential equation governing the spatial amplitude dynamics.
For standing waves (for which $\theta (x)=0$), this equation is
\begin{equation}
	\frac{d^{2}R}{dx^{2}}-\varepsilon \frac{dR}{dx}\sin (2\kappa x)+ \left[ 2
	\tilde{\mu}-\varepsilon \kappa \cos (2\kappa x)-\frac{3}{8}\varepsilon
	^{2}\cos (4\kappa x)-2V(x)\right] R -2R^{3}=0\,,  \label{nonince}
\end{equation}
where $2\tilde{\mu}=2\mu +(3/8)\varepsilon ^{2}$. Equation (\ref{nonince}),
known as a \textit{nonlinear generalized Ince equation} \cite{ngince}, is
reminiscent of the nonlinear Mathieu equation, but with the parametric force
acting on both $R$ and $R^{\prime }$.  In the linear limit, one can study Bloch waves by applying Floquet theory to Eq.~(\ref{nonince}).  In particular, one can employ the method of harmonic balance \cite{675}, in which one inserts a Fourier series expansion into the (linear) generalized Ince equation and studies the resulting infinite set of coupled linear algebraic equations satisfied by the Fourier coefficients.

\subsection{Small-Amplitude Solutions}

Let us now consider Eq. (\ref{nonince}) in the absence of the external
trapping potential $V(x)$. As shown by Eqs.~(\ref{motion}) and (\ref{effol}%
), the effective superlattice potential $\hat{V}_{\mathrm{eff}}(x)$ is
already a confining potential for the condensate. Seeking 
small-amplitude solutions, we employ the scaling $R\equiv \sqrt{\varepsilon
/2}s$, which transforms Eq.~(\ref{nonince}) into
\begin{equation}
	\frac{d^{2}s}{dx^{2}}-\varepsilon \sin (2\kappa x)\frac{ds}{dx}+\left[
	\delta ^{2}-\varepsilon \kappa \cos (2\kappa x)-\frac{3}{8}\varepsilon
	^{2}\cos (4\kappa x)\right] s-\varepsilon s^{3}=0\,,  \label{small}
\end{equation}
where $\delta^{2}\equiv 2\tilde{\mu}$ (implying that $\tilde{\mu}$ is
positive). The alternative scaling $R \equiv \varepsilon w/\sqrt{2}$ would
produce a nonlinearity of size $O(\varepsilon^{2})$ rather than $%
O(\varepsilon )$ in Eq.~(\ref{small}) and would lead one even deeper into
the small-amplitude regime.

Equation (\ref{small}) is of the form
\begin{equation}
	s^{\prime \prime}+\delta^{2}s=\varepsilon F_{1}(s,s^{\prime },x)+\varepsilon
	^{2}F_{2}(s,s^{\prime },x)\,,  \label{av}
\end{equation}
where
\begin{align}
	F_{1}(s,s^{\prime },x) &\equiv \kappa s\cos (2\kappa x)+s^{\prime }\sin
(2\kappa x)+s^{3}\,,  \notag \\
	F_{2}(s,s^{\prime },x) &\equiv \frac{3s}{8}\cos (4\kappa x)\,.  \label{force}
\end{align}
As in Ref.~\cite{map4}, we consider situations near the 2:1 subharmonic
resonance, for which $\delta =\pm \kappa $ in Eq.~(\ref{av}). 
Assuming a small ``detuning'' from the exact resonance, we introduce 
the expansion
\begin{equation}
	\delta =\kappa +\varepsilon \delta _{1}+\varepsilon
	^{2}\delta_{2}+O(\varepsilon ^{3})\,,  \label{delta}
\end{equation}
which we insert in Eq.~(\ref{av}) to obtain
\begin{equation}
	s^{\prime \prime }+\kappa ^{2}s=\varepsilon G_{1}(s,s^{\prime},x)
	+\varepsilon ^{2}G_{2}(s,s^{\prime },x)+O\left( \varepsilon ^{3}\right) \,,
\label{av2}
\end{equation}
where
\begin{align}
	G_{1}(s,s^{\prime },x)& \equiv -2\delta _{1}\kappa s +F_{1}(s,s^{\prime
},x)\,,  \notag \\
	G_{2}(s,s^{\prime },x)& \equiv \left(-\delta _{1}^{2}-2\delta _{2}\kappa) s
+ F_{2}(s,s^{\prime },x\right)\,.  \label{force2}
\end{align}

When $\varepsilon = 0$, Eq.~(\ref{av2}) has the solution
\begin{equation}
	s=\rho \cos (\kappa x+\phi )\,,  \label{base}
\end{equation}
with first derivative $s^{\prime }=-\kappa \rho \sin (\kappa x+\phi )$. We
now use the solution (\ref{base}) and its derivative as a starting point to
apply the method of variation of parameters to equation (\ref{av2}). We
therefore seek a solution to Eq.~(\ref{av2}) of the form
\begin{equation}
	s(x)=\rho (x)\cos (\kappa x+\phi (x))\,, \quad s^{\prime }(x)=-\kappa \rho
	(x)\sin (\kappa x+\phi (x))\,,  \label{vp}
\end{equation}
where $\rho(x)$ and $\phi(x)$ are slowly varying functions. Differentiating
the expression for $s(x)$ in Eq.~(\ref{vp}), we enforce the following
consistency condition with the expression for $s^{\prime }(x)$ in (\ref{vp}%
):
\begin{eqnarray}
	\rho ^{\prime } \cos (\kappa x+\phi) = \rho \phi^{\prime } \sin(\kappa
x+\phi)\,.  \label{consistency}
\end{eqnarray}

Subsequently inserting the result in Eq.~(\ref{av2}) yields
\begin{eqnarray}
	\rho ^{\prime }=&-&\frac{\varepsilon }{\kappa }\sin (\kappa x+\phi)G_{1}
\left( \rho \cos (\kappa x+\phi ),-\kappa \rho \sin (\kappa x+\phi ),x\right)
\notag \\
	&-&\frac{\varepsilon ^{2}}{\kappa }\sin (\kappa x+\phi )G_{2}\left( \rho
\cos (\kappa x+\phi ),-\kappa \rho \sin (\kappa x+\phi ),x\right)\,,
	\label{rhophi1}
\end{eqnarray}
%
\begin{eqnarray}
	\phi ^{\prime } =&-&\frac{\varepsilon }{\kappa \rho }\cos (\kappa x+\phi
)G_{1}\left( \rho \cos (\kappa x+\phi ),-\kappa \rho \sin (\kappa
x+\phi),x\right)  \notag \\
	&-&\frac{\varepsilon ^{2}}{\kappa \rho }\cos (\kappa x+\phi )G_{2}\left(
\rho \cos (\kappa x+\phi ),-\kappa \rho \sin (\kappa x+\phi),x\right)\,.
	\label{rhophi2}
\end{eqnarray}

Our objective is to isolate the components of $\rho(x)$ and $\phi (x)$ that
vary slowly as compared to the fast oscillations of $\cos(\kappa x)$ and $\sin (\kappa x) $ and to derive averaged equations governing the dynamics of these parts.
To commence averaging, we decompose $\rho$ and $\phi$ into a sum of the
slowly varying parts and small, rapidly oscillating corrections. That is,
\begin{align}
	\rho & =c+\varepsilon w_{1}(c,\varphi ,x)+\varepsilon ^{2}v_{1}(c,\varphi
,x)+O(\varepsilon ^{3})\,,  \notag \\
	\phi & =\varphi +\varepsilon w_{2}(c,\varphi ,x)+\varepsilon
^{2}v_{2}(c,\varphi ,x)+O(\varepsilon ^{3})\,.  \label{expand}
\end{align}
We then substitute Eqs.~(\ref{expand}) into Eqs.~(\ref{rhophi1})-(\ref%
{rhophi2}) and Taylor-expand to obtain
\begin{eqnarray}
	c^{\prime }&=&\varepsilon \left[ -\frac{\partial w_{1}}{\partial x}-\frac{1}{
\kappa }\sin (\kappa x+\varphi )G_{1}(c\cos (\kappa x+\varphi ),-\kappa
c\sin (\kappa x+\varphi ),x)\right]  \notag \\
	&+&\varepsilon ^{2}\left[ -\frac{\partial v_{1}}{\partial x}+L_{1}(c,\varphi
,x)\right] +O(\varepsilon ^{3})\,,  \label{c'phi'1}
\end{eqnarray}
%
\begin{eqnarray}
	\varphi ^{\prime }&=&\varepsilon \left[ -\frac{\partial w_{2}}{\partial x}-
\frac{1}{\kappa c} \cos (\kappa x+\varphi )G_{1}(c\cos (\kappa x+\varphi
),-\kappa c\sin (\kappa x+\varphi ),x)\right]  \notag \\
	&+&\varepsilon ^{2}\left[ -\frac{ \partial v_{2}}{\partial x}%
+L_{2}(c,\varphi ,x)\right] +O(\varepsilon ^{3})\,.  \label{c'phi'2}
\end{eqnarray}
The functions $w_{1}$ and $w_{2}$ in Eqs.~(\ref{expand})-(\ref{c'phi'2})
should be chosen so as to eliminate all the rapidly oscillating terms at
order $O(\varepsilon )$. At $O(\varepsilon^2)$, the functions $v_{1}$ and $%
v_{2}$ are similarly chosen. The terms $L_{1}(c,\varphi ,x)$ and $%
L_{2}(c,\varphi ,x)$ in Eqs. (\ref{c'phi'1})-(\ref{c'phi'2}) depend on $%
w_{1} $, $w_{2}$, $G_{1}$, and $G_{2}$. We provide expressions for $w_{1}$, $%
w_{2}$ and $L_{1}$, $L_{2}$ in the Appendix. (Note that expressions for $%
v_{1}(x)$ and $v_2(x)$ are not needed until one tackles the third-order
corrections, so we do not display them.) 
The functions $G_{1}$ and $G_{2}$ are expressed in terms of $c$ and $\varphi$
as follows:
\begin{align*}
	G_{1}(c,\varphi ,x) &= - 2\delta _{1}\kappa c\cos (\kappa x+\varphi) +\kappa
c\cos (\kappa x+\varphi )\cos (2\kappa x) \\ 
&- \kappa c\sin (\kappa x+\varphi
)\sin (2\kappa x)+c^{3}\cos ^{3}(\kappa x+\varphi )\,, \\
	G_{2}(c,\varphi ,x)&= \left(-\delta _{1}^{2}-2\delta _{2}\kappa\right) c\cos
(\kappa x+\varphi) +\frac{3}{8} c\cos (\kappa x+\varphi )\cos (4\kappa x)\,. \nonumber
\end{align*}

To first order, the slow evolution equations are
\begin{equation}
	c^{\prime }=O\left( \varepsilon ^{2}\right) \,,~\varphi ^{\prime
	}=\varepsilon \left( \delta _{1}-\frac{3c^{2}}{8\kappa }\right)
	+O(\varepsilon ^{2})\,,  \label{first-order}
\end{equation}
generating a circle of nonzero equilibria with amplitude $c_{\ast }=\sqrt{
8\delta _{1}\kappa /3}$. The generating functions $w_{1}(x)$ and $w_{2}(x)$
that yield these equations are shown in the Appendix. This process results
in the amplitude function
\begin{equation}
	R(x)=\sqrt{\frac{\varepsilon }{2}}c_{\ast }\cos (\kappa x+\varphi _{\ast
})\equiv v(x,t=0)\,,  \label{rr}
\end{equation}
where the phase shift $\varphi _{\ast }$ is arbitrary. The corresponding MAW
is given by $v(x,t)=R(x)\exp (-i\mu t)$, as per Eq.~(\ref{maw2}).

\begin{figure}[tbp]
\centerline{
\includegraphics[width = 0.4\textwidth]{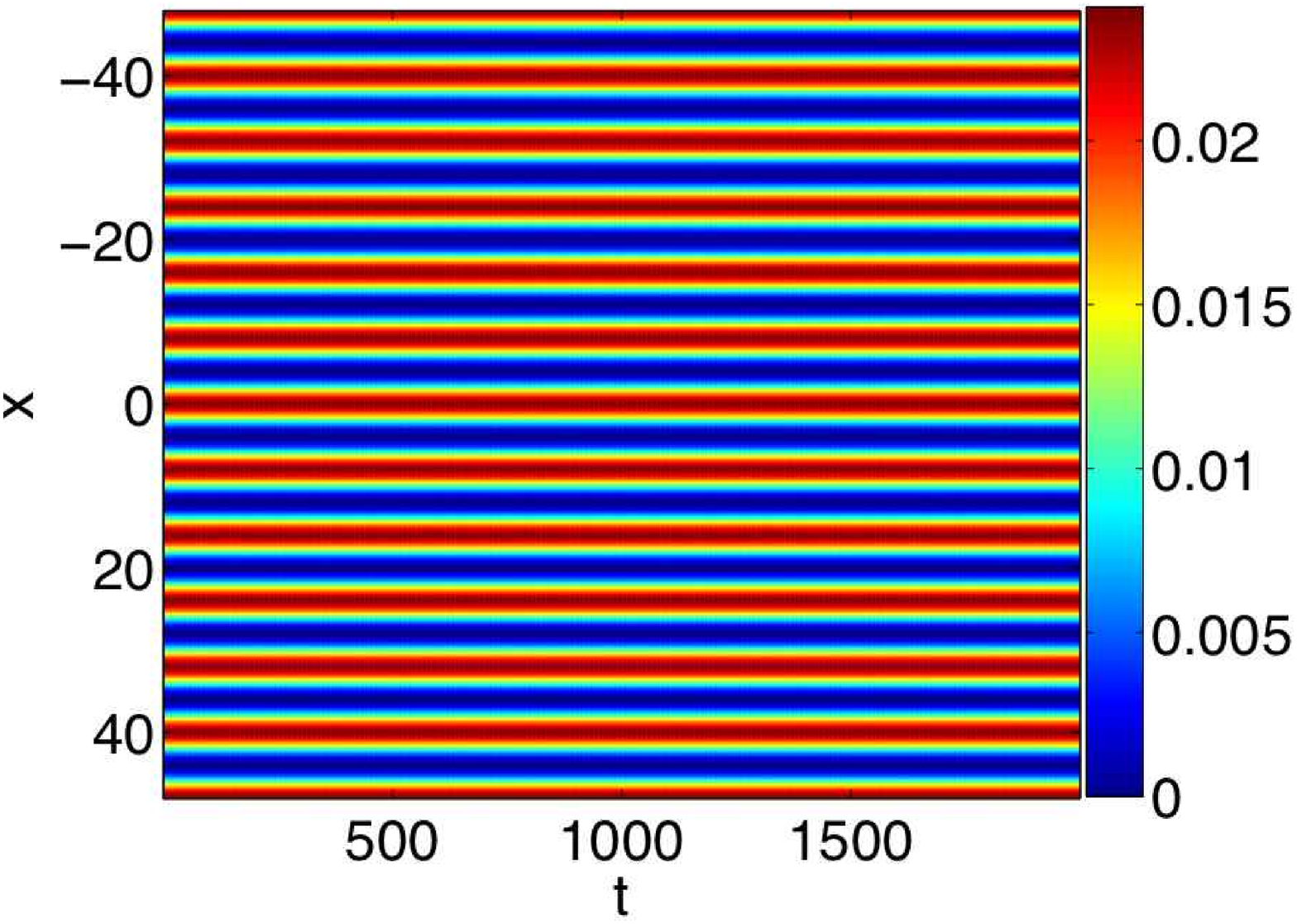}
\hspace{.2 cm}
\includegraphics[width = 0.4\textwidth]{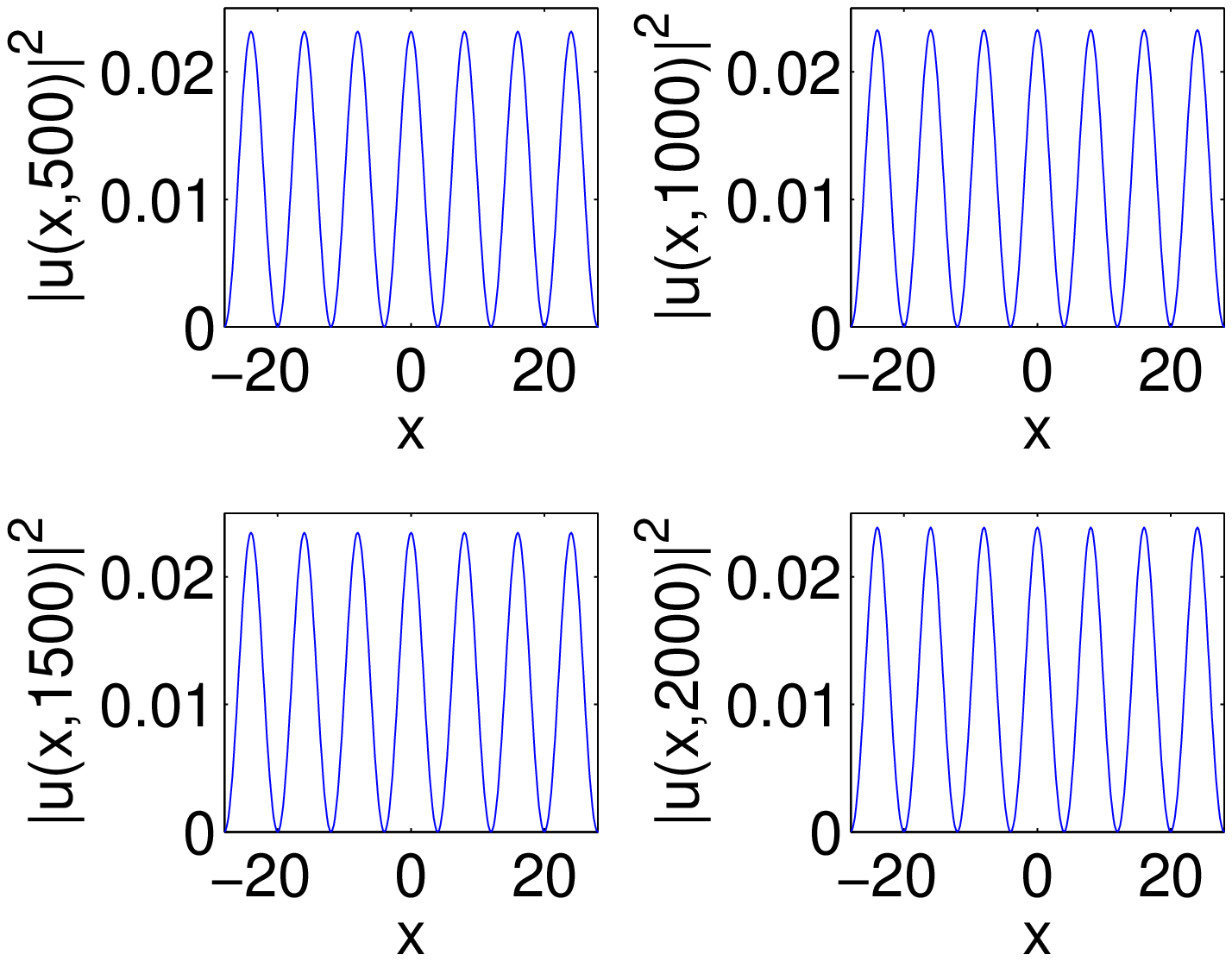}}
\caption{(Color online) Dynamical evolution of the sinusoidal MAW of Eq.~(%
\protect\ref{rr}) with parameter values $\protect\phi_{\star}=0$ for $g_0=2$%
, $\protect\kappa=\protect\pi/8$, $V_0=0.15$, $\protect\delta_{1}=3$, and $%
\protect\delta _{2}=1$. The left panel shows the space-time contour plot of
the square modulus (density) $|u|^2$ of the solution, and the right panels
show four snapshots (spatial profiles) of the spatio-temporal evolution. The
dynamics illustrate an apparent stability of the solution up through at
least $t = 2000$.}
\label{original1}
\end{figure}

\begin{figure}[tbp]
\centerline{
\includegraphics[width =0.4\textwidth]{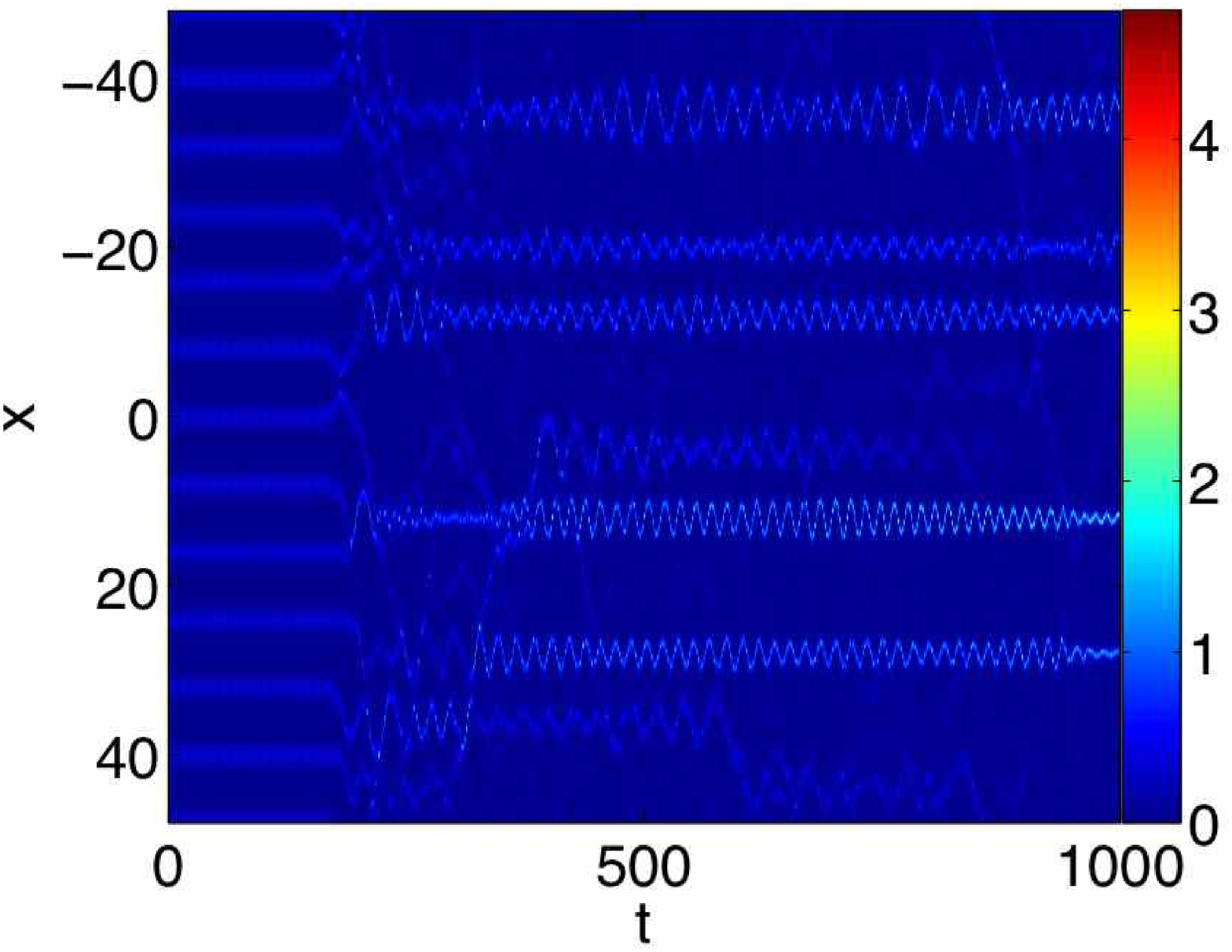} \hspace{.2 cm}
\includegraphics[width = 0.4\textwidth]{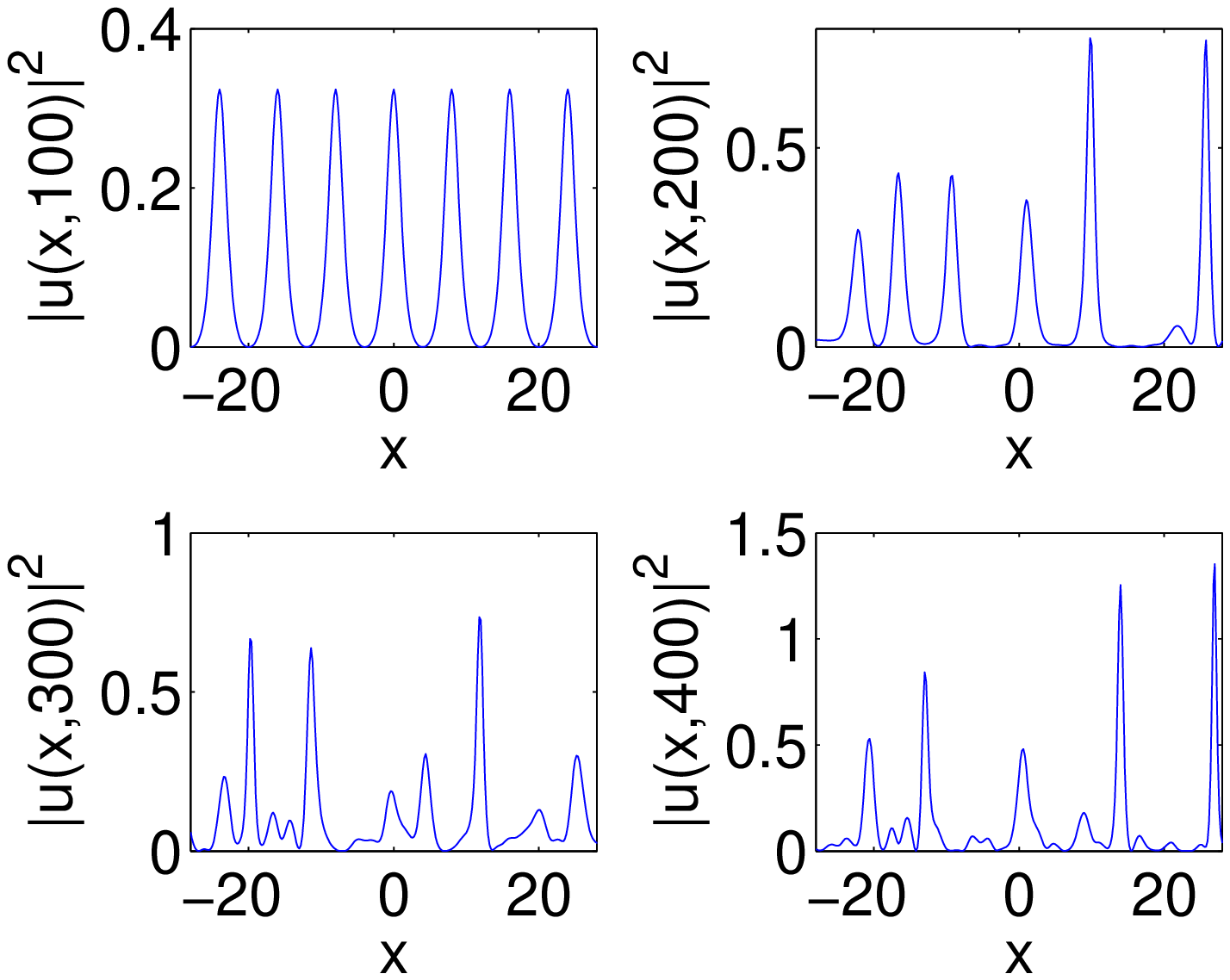}}
\caption{(Color online) Same as Fig.~\protect\ref{original1}, but for $%
V_0=1.5$. As early as $t=200$, the MAW solution starts breaking up into
solitary filaments that appear to be localized around the maxima of the
nonlinear optical lattice (see the discussion in the text).}
\label{original2}
\end{figure}

Returning to the original GP equation (in the absence of an external
potential), we obtain the MAW $u(x,t)=v(x,t)/\sqrt{g(x,t)}$. We examine its
dynamics and stability with direct simulations of Eq.~(\ref{nls3}) with $%
V(x)=0$. Results of the simulations for $g_{0}=2$, $V_{0}=0.15$, $\kappa=\pi/8$ (so that $\varepsilon \approx 0.0295$), $\delta_{1}=3$, $\delta _{2}=1$,
and $\varphi_{\ast }=0$ are shown in Fig.~\ref{original1}. These
parameter values correspond to a $^{87}$Rb (or a $^{23}$Na) condensate in a
trap with transverse confining frequency $\omega_{\perp}=2 \pi \times 1000$%
Hz that contains $N \approx 10^3$ atoms with a peak 1D density $n_0 = 10^8$ m%
$^{-1}$. Here, the spatial unit (i.e., the healing length) is $\xi = 0.3
\,\mu$m ($\xi = 0.9\, \mu$m), and the temporal unit is $\xi/c = 0.14$ ms ($%
\xi/c = 0.3$ ms). Figure ~\ref{original1} shows that the MAW appears to be stable for long times as a solution of the original GP equation. For larger $\varepsilon$, however, the wave breaks up
into solitary filaments (that appear to be localized around the maxima of
the nonlinear optical lattice; see the discussion below), as indicated in
Fig.~\ref{original2} (for $V_0 = 1.5$). In fact, as we will discuss below,
the apparently stable MAWs obtained for small $V_0$ are actually weakly
unstable, although their lifetimes are very long. The normalized time $t = 2000$
amounts to $280$ ms for a Rubidium (and $600$ ms for a Sodium) BEC in real
time, suggesting that these MAWs can nevertheless be observed in experiments.

\begin{figure}[tbp]
\centerline{
\includegraphics[width = 0.4\textwidth]{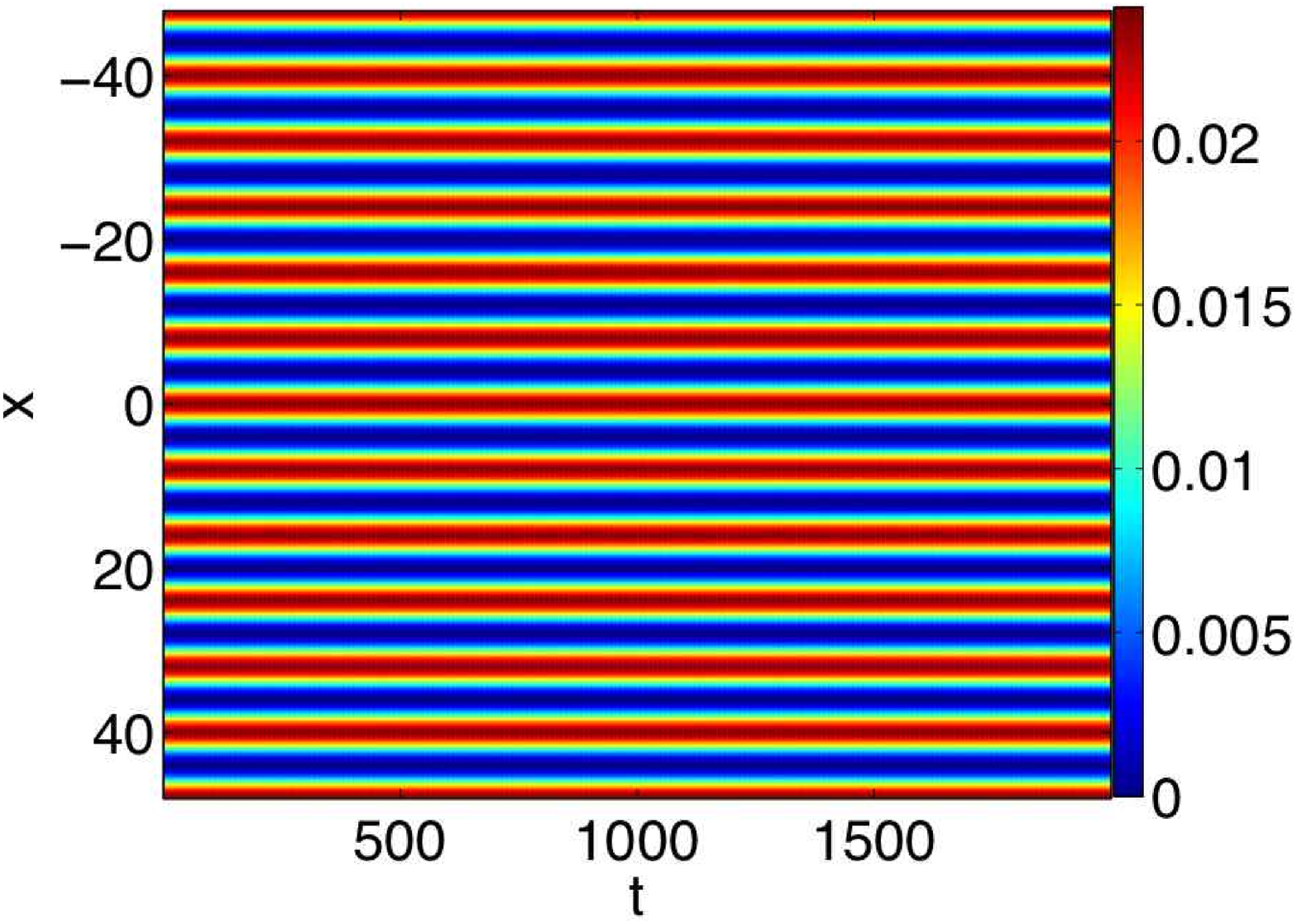}
\hspace{.2 cm}
\includegraphics[width = 0.4\textwidth]{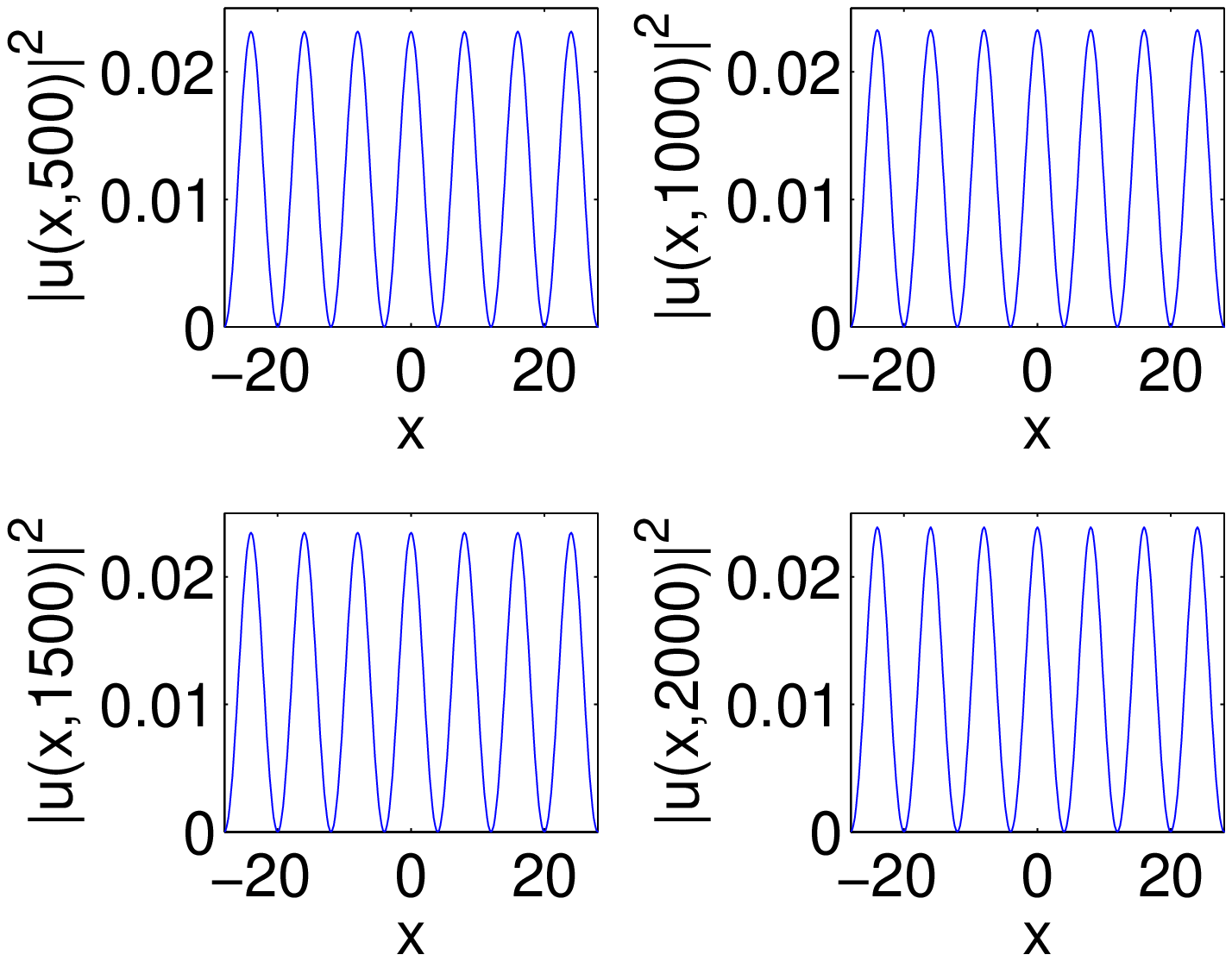}}
\caption{(Color online) Same as Fig.~\protect\ref{original1} (and also with $%
V_0=0.15$), but for the refined prediction of the initial MAW solution based
on Eq.~(\protect\ref{ww}). The solution also appears to be dynamically
stable.}
\label{with1}
\end{figure}

\begin{figure}[tbp]
\centerline{\includegraphics[width =
0.4\textwidth]{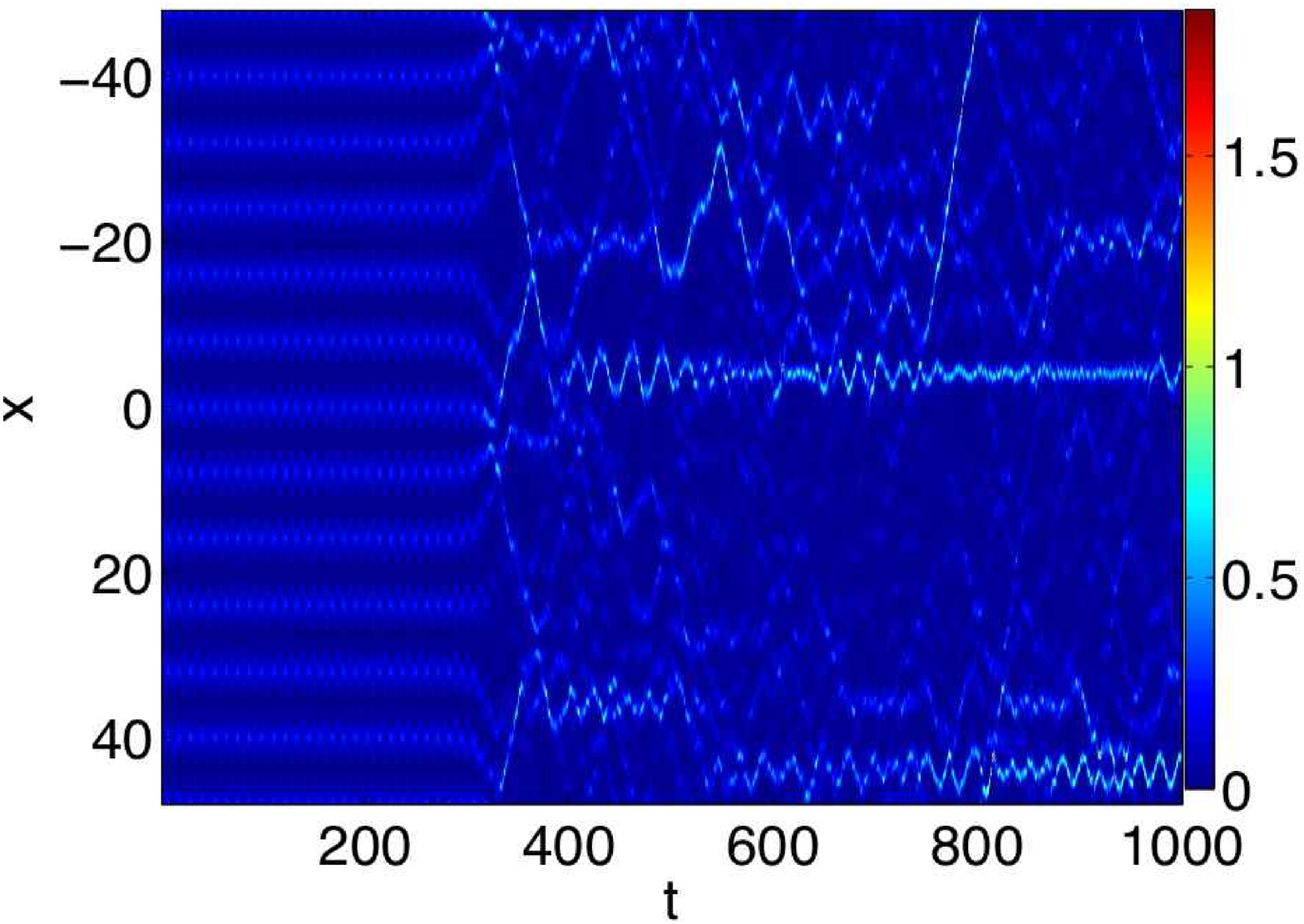} \hspace{.2 cm}
\includegraphics[width =
0.4\textwidth]{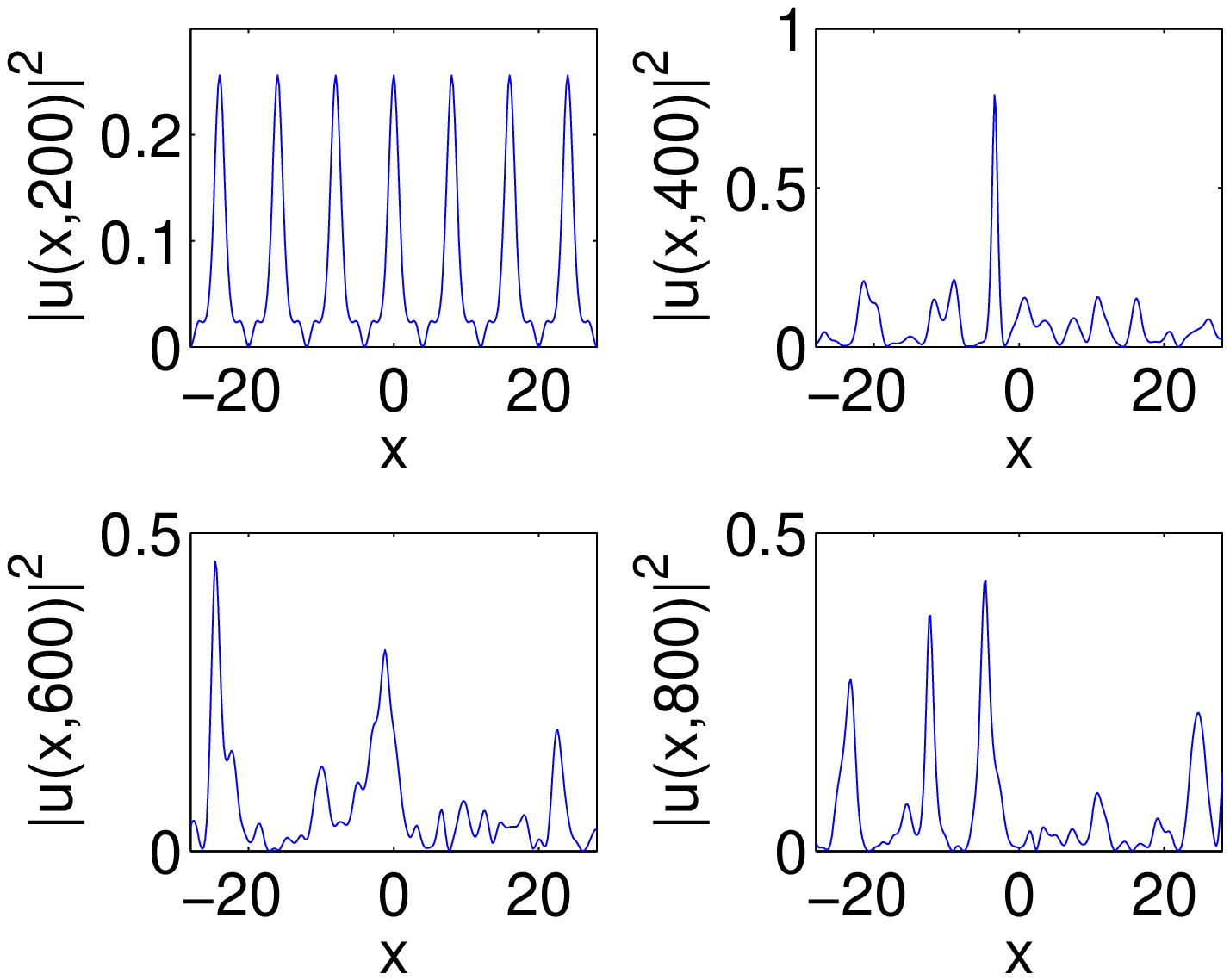}}
\caption{(Color online) Same as Fig.~\protect\ref{original2}, but with the
MAW of Eq.~(\protect\ref{ww}) used as the initial condition for the
time-evolution of the GP equation. Observe that the refined initial
condition leads to a delayed initiation of the instability (which now begins
at about $t \approx 300$).}
\label{with2}
\end{figure}

One can also incorporate the generating functions into the MAWs and examine
the stability properties of the resulting refined MAWs using direct
numerical simulations. These MAWs are given by
\begin{equation}
	R(x)=\sqrt{\frac{\varepsilon }{2}}(c_{\ast }+\varepsilon w_{1})\cos (\kappa
	x+\varphi _{\ast }+\varepsilon w_{2})\equiv v(x,t=0)  \label{ww}
\end{equation}
instead of Eq.~(\ref{rr}). We plot the corresponding space-time 
plots and spatial profiles at various time 
instants for $V_{0}=0.15$ in Fig.~\ref{with1} and for $V_{0}=1.5$ in Fig.~%
\ref{with2}. All other parameters are the same as before. It is readily
observed that the instability in Fig.~\ref{with2} 
sets in later (at $t \approx 300$) than that in Fig.~\ref{original2} (at $t
\approx 200$), which was obtained with the same parameter values using the
sinusoidal MAW of Eq.~(\ref{rr}).

The MAWs in Eqs.~(\ref{rr}) and (\ref{ww}) were determined to order $%
O(\varepsilon)$, so they ignore the effects of the second-harmonic component
of the lattice, which arise at $O(\varepsilon^{2})$. To estimate the
influence of this component, we used the same MAWs as initial conditions in
direct simulations of the original GP equations in the presence of an
additional external optical lattice (OL) potential, $V(x)=-(3/16)\varepsilon
^{2}\cos (4\kappa x)$ that exactly cancels out the second-harmonic component
of the lattice. While differences between the results of the simulations
with and without the compensating OL are not apparent in comparing
space-time plots side-by-side, one can observe the slow development of small
discrepancies by examining the time-evolution of the absolute value of their difference.

To second order, the equations of slow evolution are
\begin{align}
	c^{\prime }& =\varepsilon ^{2}\left[ \frac{c}{4\kappa }\left( \delta _{1}-
\frac{c^{2}}{8\kappa }\right) \sin (2\varphi )\right] +O(\varepsilon ^{3})\,,
\notag \\
	\varphi ^{\prime }& =\varepsilon \left( \delta _{1}-\frac{3c^{2}}{8\kappa }
\right) +\varepsilon ^{2}\left[ \delta _{2}-\frac{1}{16\kappa }+\frac{
3c^{2}\delta _{1}}{4\kappa ^{2}}-\frac{51c^{4}}{256\kappa ^{3}}+\frac{1}{
4\kappa }\left( \delta _{1}-\frac{c^{2}}{4\kappa }\right) \cos (2\varphi ) %
\right] +O(\varepsilon ^{3})\,.  \label{second}
\end{align}
In contrast to the first-order equations (\ref{first-order}), the
equilibrium points of Eqs.~(\ref{second}) depend on $\varepsilon$,
which is typical in second-order averaging. In studying the dynamics of
slow-flow equations produced by such a procedure, one obtains, in general, a
complicated bifurcation problem, which can be investigated by taking $%
\varepsilon $ small but fixed (see, e.g., Ref.~\cite{ngmat}). For our
purposes, we simply note that the only difference between Eqs.~(\ref{second}%
) and the slow equations one would obtain at second order without the $%
O(\varepsilon ^{2})$ lattice term in Eq.~(\ref{nonince}) amounts to constant
terms at the same order [they result directly from the detuning of $\delta $%
; see Eq.~(\ref{delta})]. Without the second-order lattice term, $\delta
_{2} $ would not be present, and there would be an additional constant term,
$-\delta _{1}^{2}/(2\kappa )$, that is cancelled out here because of the
extra harmonic. To study the effects of the second-order lattice
systematically, one can examine the dynamics starting with the alternative
scaling $R \equiv \varepsilon w/\sqrt{2}$ (rather than $R\equiv \sqrt{%
\varepsilon /2}s$, as adopted above), in which case the effect of the
nonlinearity also emerges at second order.

In the absence of resonances, second-order averaging yields slow dynamical
equations for $(c,\varphi )$ with $c^{\prime }=O\left( \varepsilon
^{3}\right) $ and $\varphi ^{\prime }<0$, so that a fixed-radius circle is
traversed in the clockwise direction. From here, one can also consider
resonances whose effects appear at $O(\varepsilon ^{2})$.

\subsection{Stability}

In Figs.~\ref{original2} and \ref{with2}, we observe that the MAW initial
conditions, derived for small $V_0$, break down rather quickly for larger $%
V_0$. In fact, there is a weak instability even for small $V_0$, although
the time it takes for the instability to set in is rather long (beyond $t =
2000$). Thus, the MAW solutions have a good chance to be observed
experimentally for sufficiently small scattering-length modulations $V_0$.  (Recall that $t=2000$ corresponds to $280$ ms for a $^{87}$Rb BEC and $600$ ms for a $^{23}$Na one.)

\begin{figure}[tbp]
\centerline{\includegraphics[width = 0.9\textwidth]{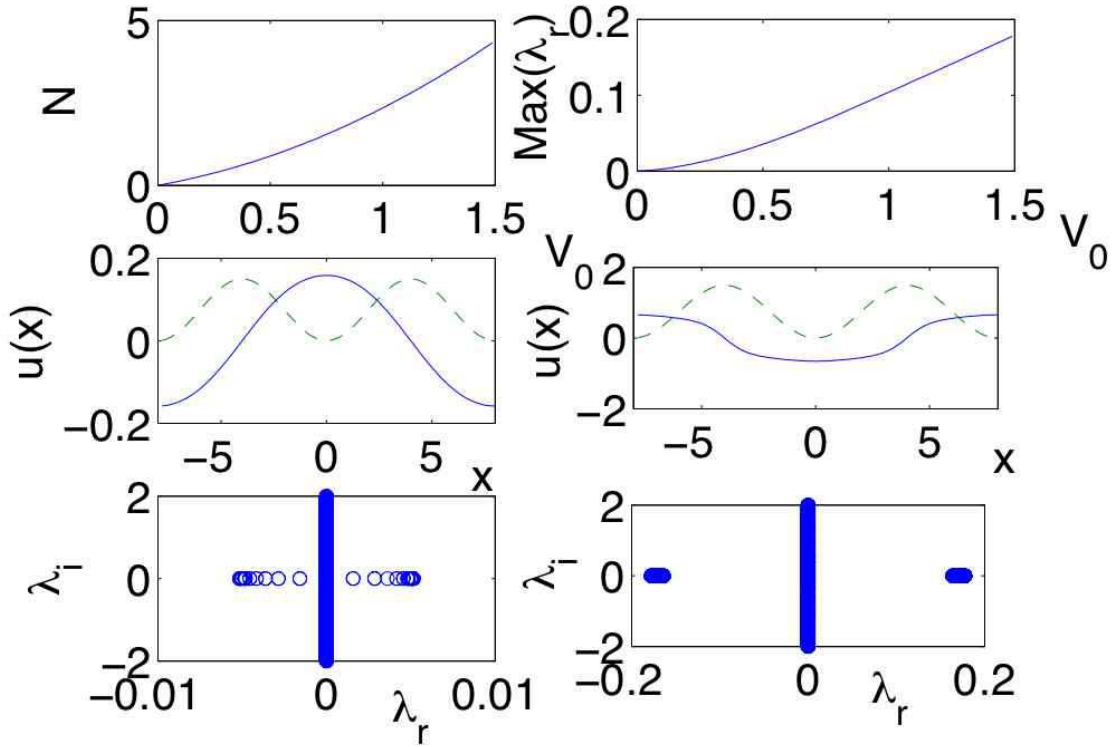}}
\caption{(Color online) Results of a fixed point iteration to identify the
solution given by Eq.~(\protect\ref{rr}) as a numerically exact stationary
state. The top left panel shows the $L^2$ norm (density) of this state in
our computational domain (containing one period of the solution), and the
top right panel shows the maximal real part of the most unstable eigenvalue
of the configuration. Observe that the configuration is \textit{always}
unstable. Hence, the apparent stability of Figs.~\protect\ref{original1} and
\protect\ref{with1} arises only because of the finite time (up to $t = 2000$%
) of the direct GP simulations. The middle left and right panels show the
solution (solid curve) and nonlinear lattice $g(x)$ (dashed curve) for $%
V_0=0.15$ and $V_0=1.5$, respectively. The bottom left and right panels show
the corresponding spectral planes of the solutions and indicate the weak
instability of the former and the strong instability of the latter.}
\label{full1}
\end{figure}

To investigate this point further, we performed fixed-point computations
using Eqs.~(\ref{rr}) and (\ref{ww}) as starting guesses in order to obtain
\textit{numerically exact} stationary states. We show the results for Eq.~(%
\ref{rr}) in Fig.~\ref{full1} using a computational domain containing one
period of the solution. In this context, we have implemented a variant of
our finite-difference method (for the spatial discretization) that
incorporates Floquet theory, as is described in, e.g., \cite{bernard}. Adding an $\exp(i \theta)$ term and its complex conjugate at the appropriate locations of our stability matrix, we fill in the bands
of continuous spectrum by varying $\theta$ within the interval $[0,2\pi)$.
We used a partition of $200$ equidistant points in $\theta$ for the results
shown here. We find that the configuration is \textit{always} unstable, even
for the small $V_0$ that appeared to be stable based on direct numerical
simulations. Hence, the apparent stability of Figs.~\ref{original1} and \ref%
{with1} arises only because of the fact that the simulation was performed
for a finite time (up to $t = 2000$). As indicated in Fig.~\ref{full1}
(where the spectral planes are plotted in the bottom panels), the
instability is weak for small $V_0$ but becomes strong for large $V_0$.

\begin{figure}[tbp]
\centerline{
(a) \includegraphics[width = 0.4\textwidth]{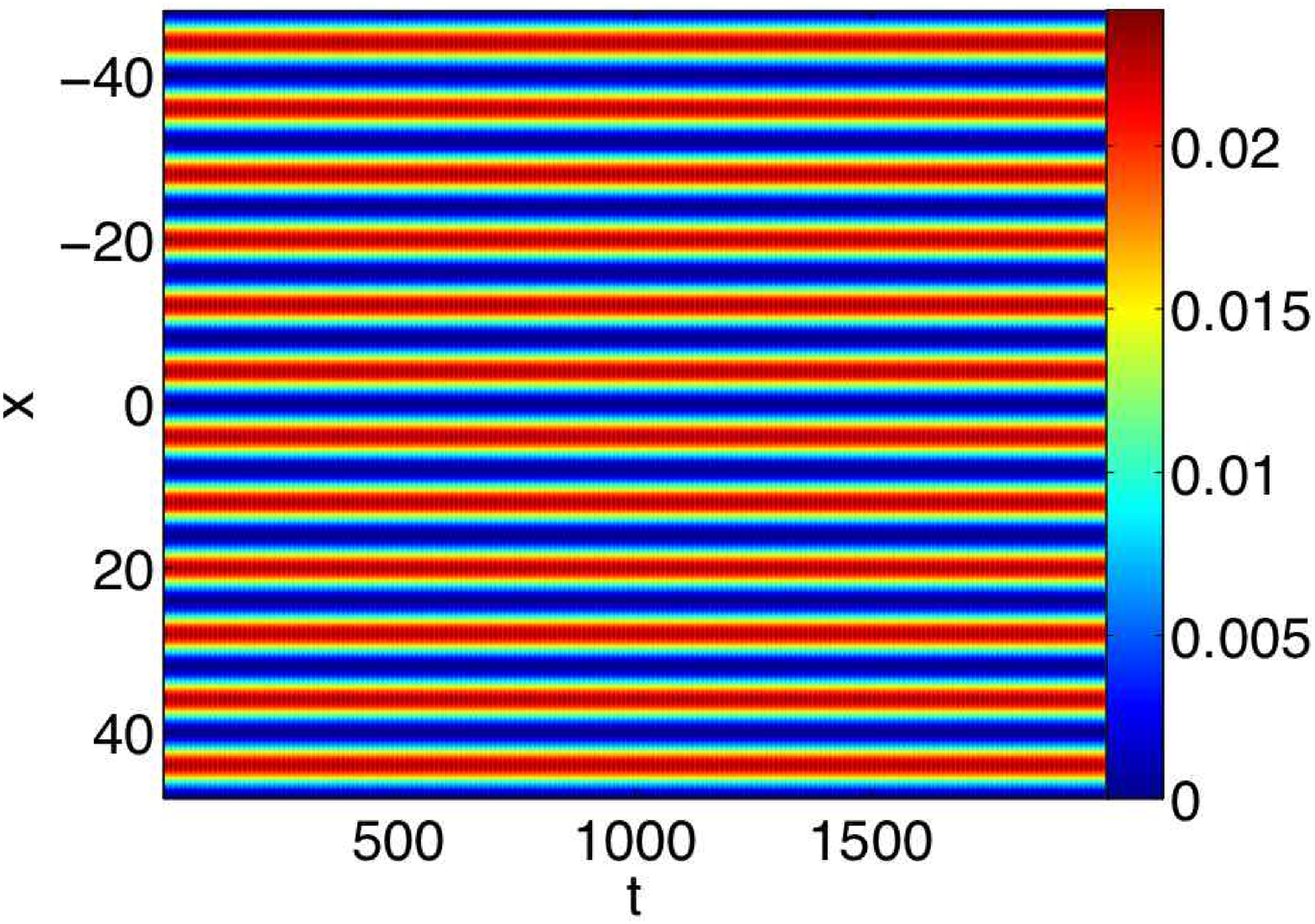}
\hspace{.2 cm}
(b) \includegraphics[width = 0.4\textwidth]{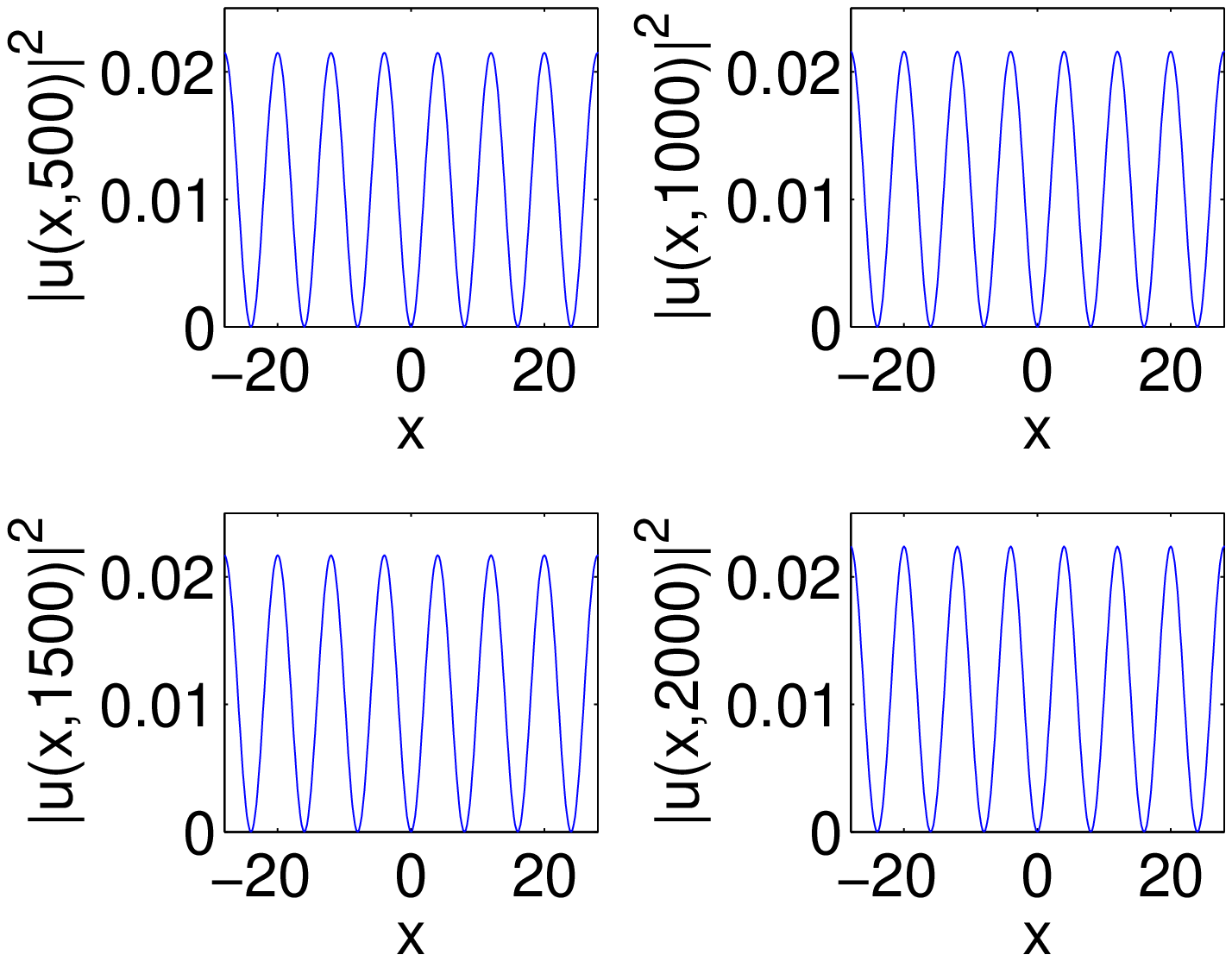}}
\centerline{
(c) \includegraphics[width = 0.4\textwidth]{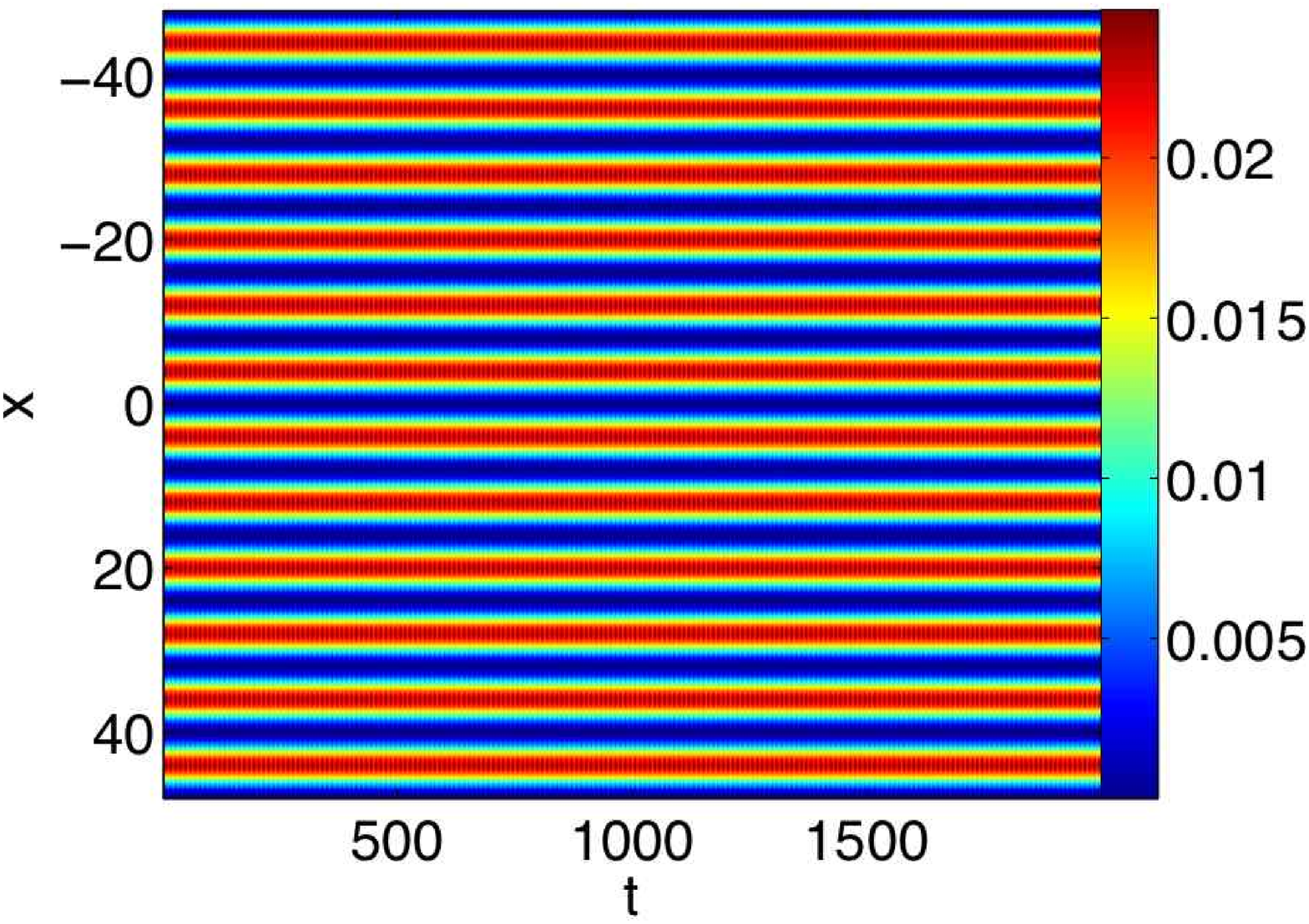}
\hspace{.2 cm}
(d) \includegraphics[width = 0.4\textwidth]{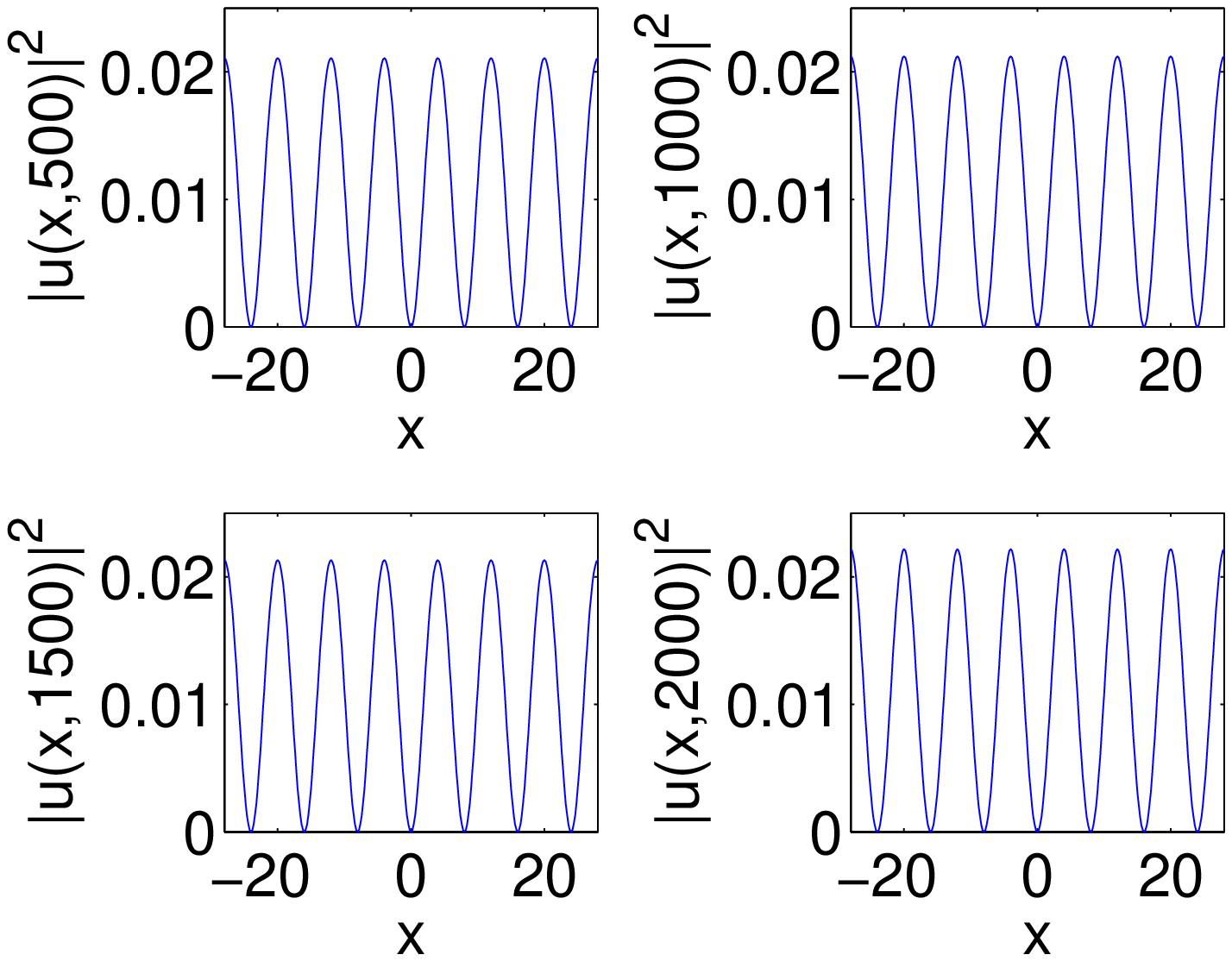}}
\caption{(Color online) Dynamical evolution of the \textit{on-site}
solutions with $\protect\phi_{\star}=\protect\pi/2$ for the initial waves of
Eq.~(\protect\ref{rr}) [panels (a) and (b)] and Eq.~(\protect\ref{ww})
[panels (c) and (d)]. In both cases, $V_0=0.15$.}
\label{onsite1}
\end{figure}

To find more stable solutions, we note that the transformation $v\equiv
\sqrt{g(x)}u$ suggests that for the nonlinear OL, solutions prefer to be centered \textit{on-site} (in contrast to what occurs for linear OLs), so that their
maxima coincide with \textit{maxima} of the lattice rather than minima. This
is not surprising, as the multiplicative terms (i.e., the ones without the $%
d/dx$) of $V_{\mathrm{eff}}(x)$ act as a regular (super)lattice after the
transformation and the minima of $g(x)$ coincide with the maxima of these
terms in $V_{\mathrm{eff}}(x)$. (We note in passing that this difference
between linear and nonlinear OLs has also recently been highlighted in \cite%
{wein06} from a spectral theory perspective for bright soliton solutions of
the NLS equation.) We observe additionally in Figs.~\ref{original2} and \ref%
{with2} that when the MAWs break up, one obtains solutions that are
localized on maxima of the nonlinear OL. We can take advantage of this
observation by using MAWs with different phases $\varphi _{\ast }$ as
initial wave functions in simulations of the GP equation (\ref{nls3}).

\begin{figure}[tbp]
\centerline{ (a) \includegraphics[width =
0.4\textwidth]{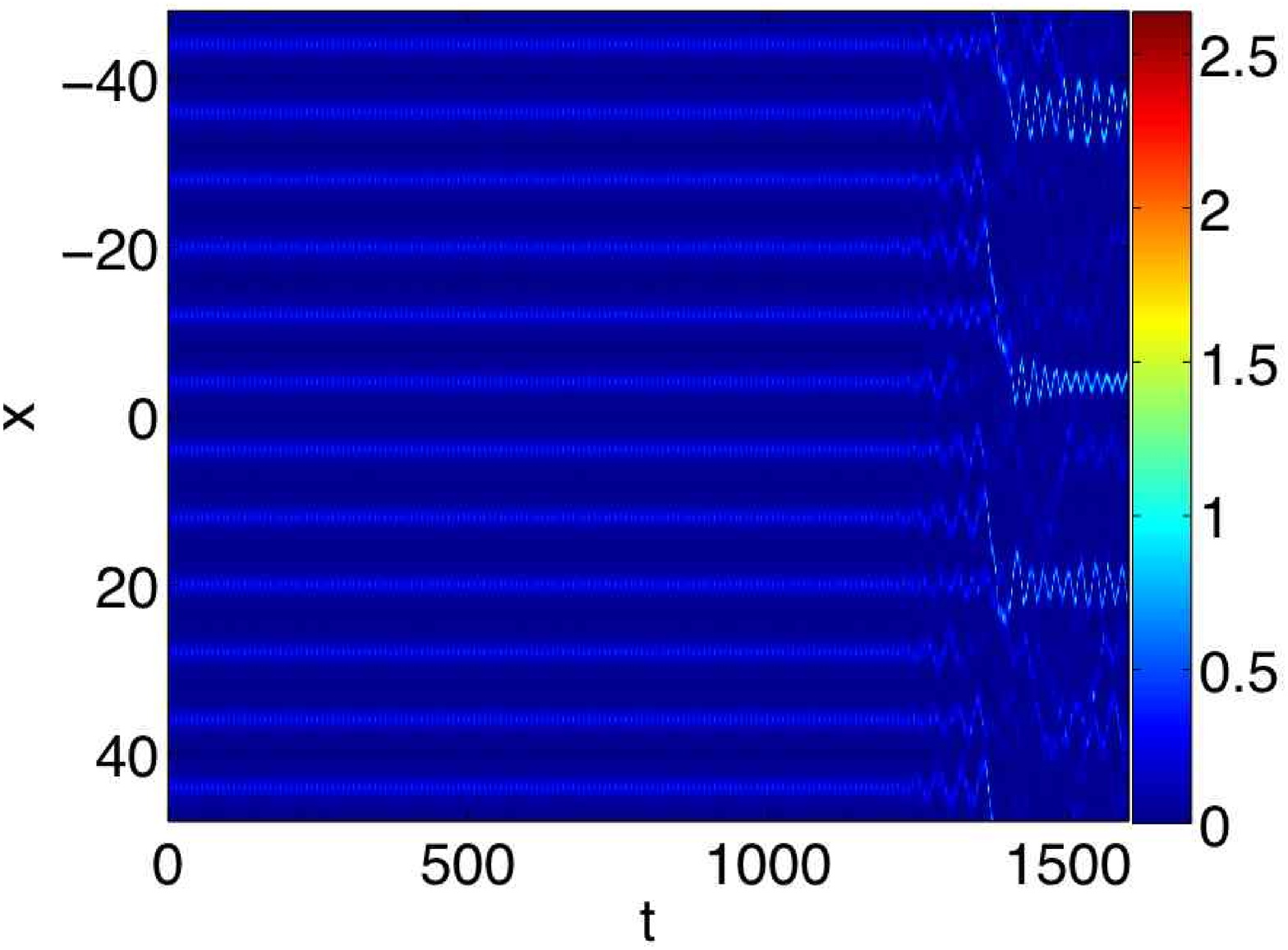} \hspace{.2 cm} (b)
\includegraphics[width =
0.4\textwidth]{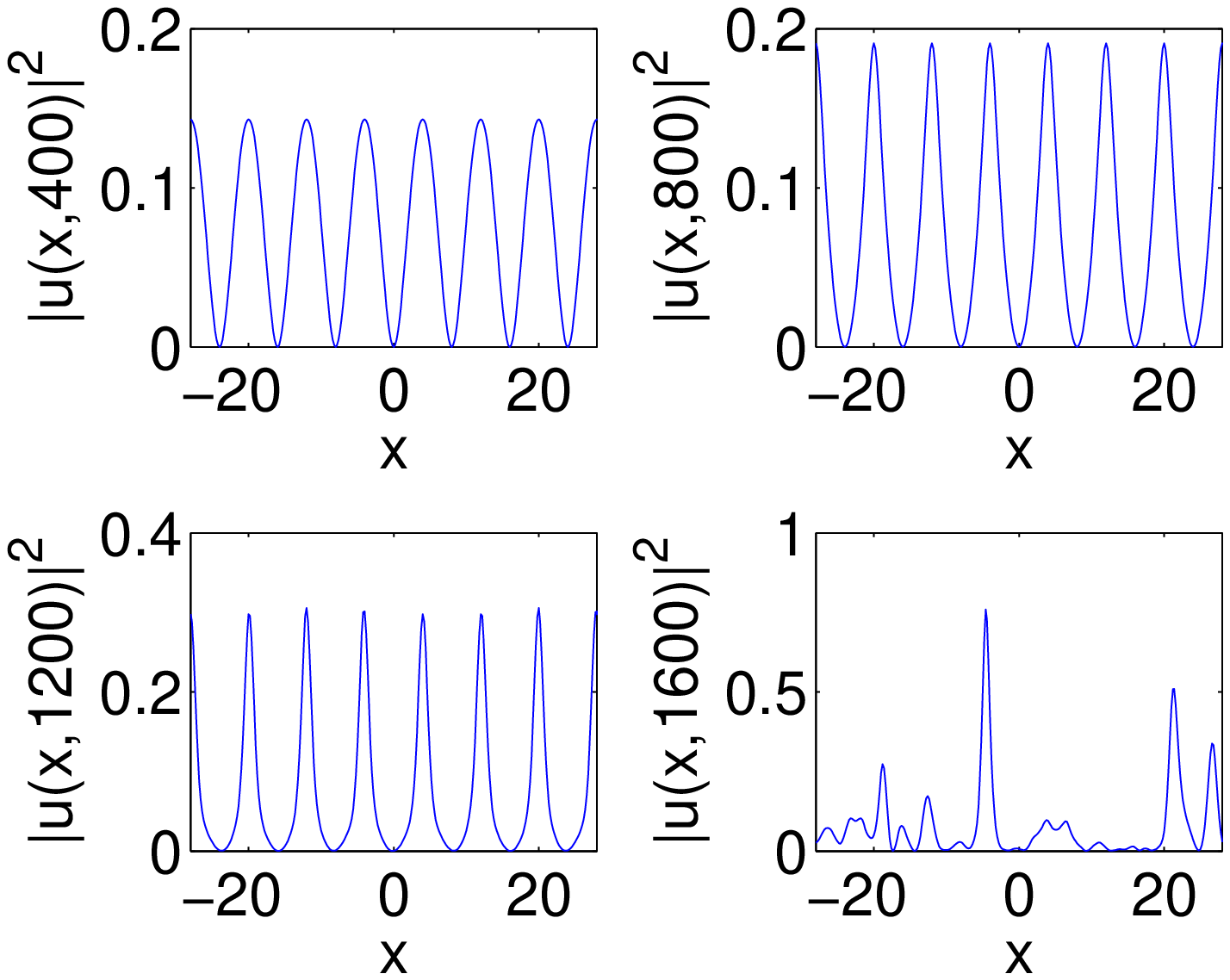}}
\centerline{ (c) \includegraphics[width =
0.4\textwidth]{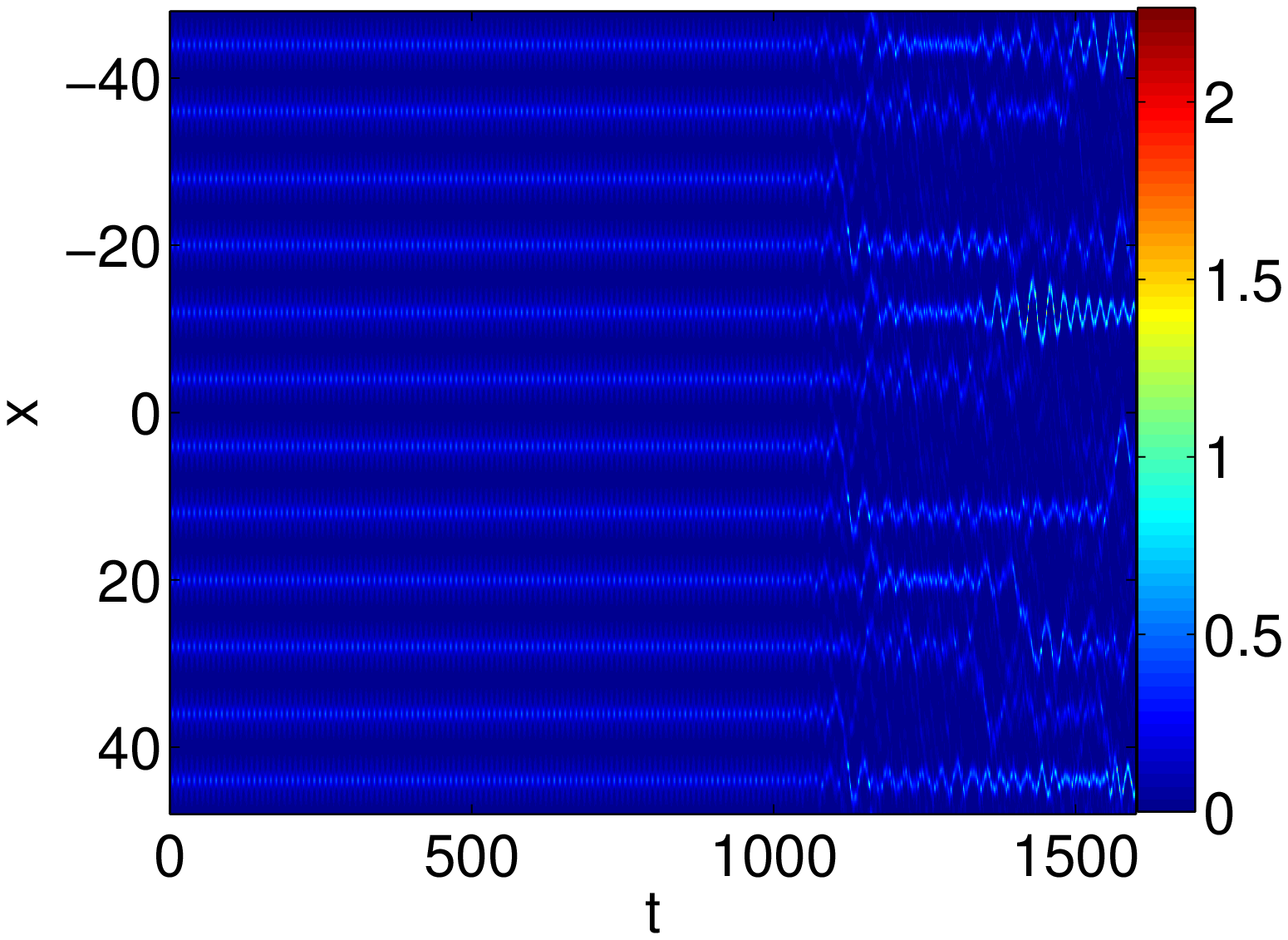} \hspace{.2 cm} (d)
\includegraphics[width =
0.4\textwidth]{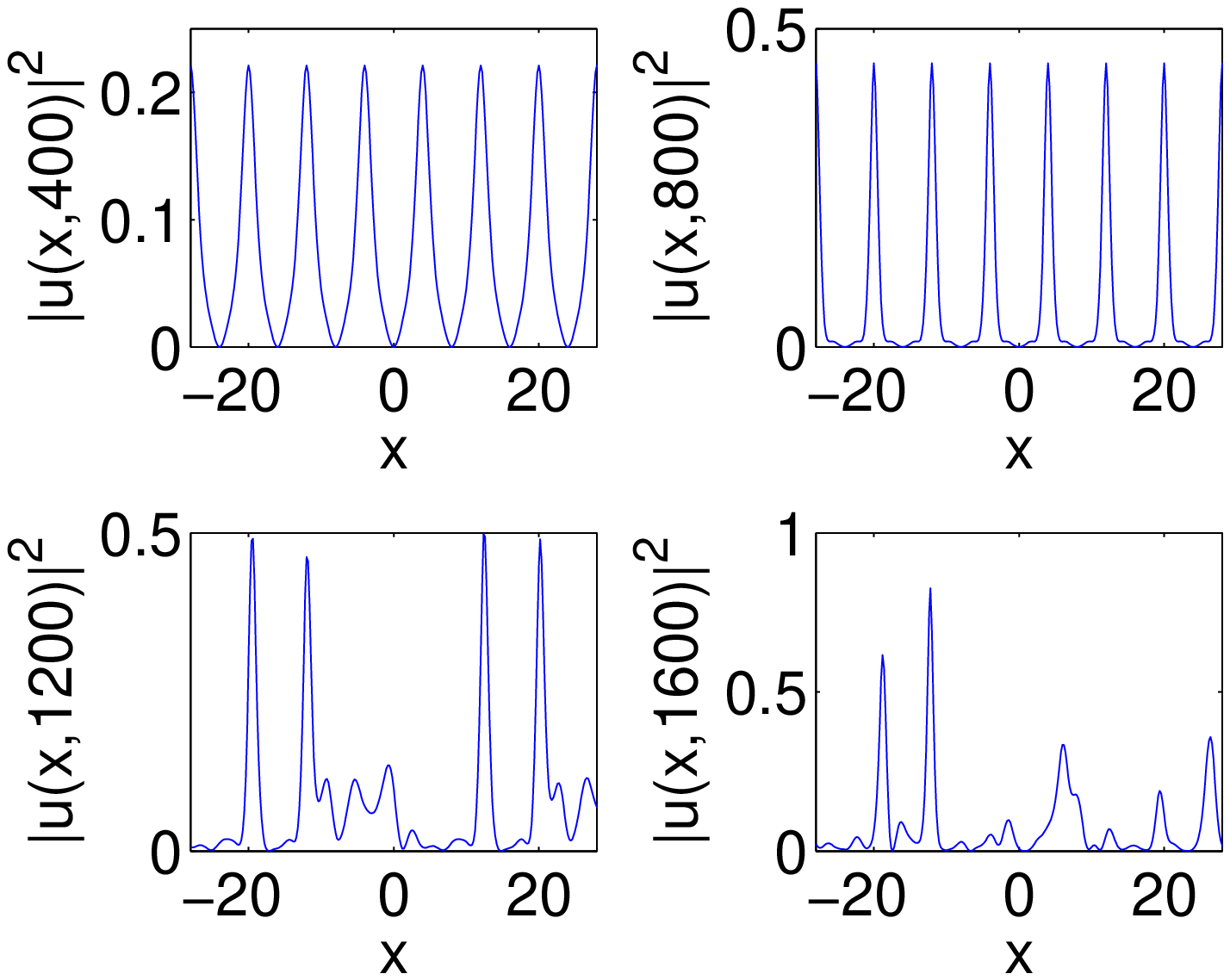}}
\caption{(Color online) Same as Fig.~\protect\ref{onsite1}, except for $%
V_0=1.5$. Observe that the solutions now become unstable for times longer
than $t=1000$, highlighting the much weaker nature of the corresponding
instability in comparison to its off-site counterpart. The fact that the
refined ansatz (\protect\ref{ww}) now seems to become unstable at an earlier
time is probably due to the significant change in the profile of the (exact)
solution for large $V_0$ (see Fig.~\protect\ref{full2}).}
\label{onsite2}
\end{figure}

\begin{figure}[tbp]
\centerline{ \includegraphics[width = 0.9\textwidth]{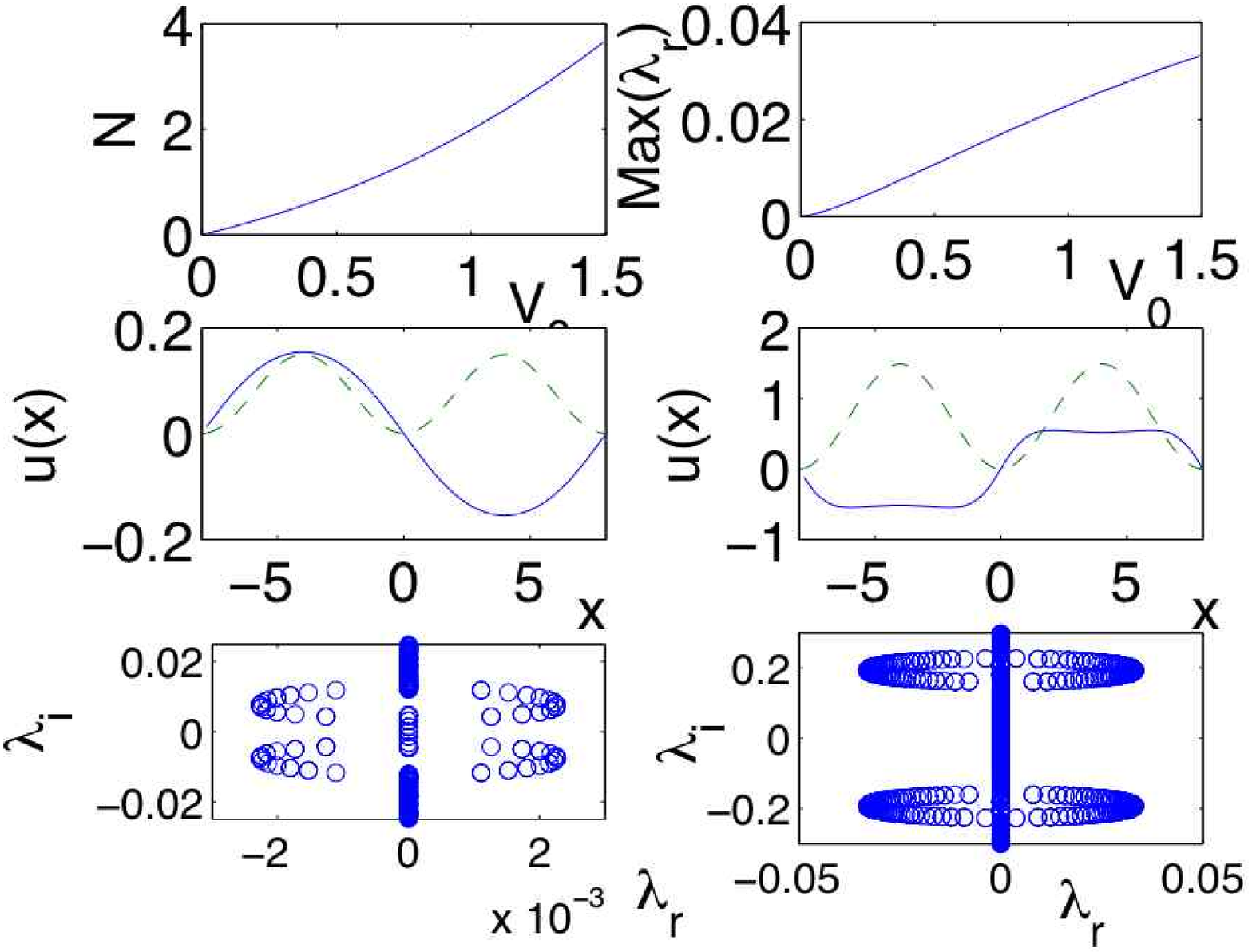}}
\caption{(Color online) Same as Fig.~\protect\ref{full1}, except for on-site
MAWs. In this case, the instability of the solutions is much weaker and is oscillatory in nature.}
\label{full2}%
\end{figure}

We show the dynamical evolution of on-site MAWs (for which $\varphi_* =
\pi/2 $) in Figs.~\ref{onsite1} and \ref{onsite2} for $V_0 = 0.15$ and $V_0
= 1.5$, respectively. As observed in Fig.~\ref{onsite2}, the solution
remains stable past $t = 1000$ instead of breaking up far earlier, as was
the case for off-site solutions. We confirm these observations on stability
using fixed-point computations of the same type as above (but now for
solutions with $\varphi_*=\pi/2$), the results of which are shown in Fig.~%
\ref{full2}. We observe that the maximum eigenvalue now has a much smaller
value and that the ensuing instabilities here are much weaker oscillatory
ones. An interesting feature that we note in passing is the formation of
``rings'' of such oscillatory instabilities (see Fig.~\ref{full2}).

The stability results discussed above can also be considered in light of the
recent work \cite{zoilong} on the development of instabilities for NLS
equations with constant nonlinearity coefficients and periodic potentials.
The authors of Ref.~\cite{zoilong} found (among other results) that
small-amplitude, periodic, standing-wave solutions corresponding to band
edges alternate in their stability, where upper band edges are
modulationally unstable in the attractive case and lower band edges are
modulationally unstable in the repulsive case. The theory in \cite%
{zoilong} is also applicable to the MAW solutions constructed above, as the
resonance relation they satisfy guarantees their spatial periodicity.

To examine the onset of modulational instabilities, we consider small
perturbations to the MAWs of the form $u_1\exp(\lambda t - i\mu t)$. [That
is, one linearizes the GP equation (\ref{nls3}) around the MAW solutions.]
Such perturbations satisfy the Bogoliubov equations,
\begin{equation}
	L_+ p = \lambda q\,, \quad L_-q = \lambda p\,,
\end{equation}
where the amplitude $u_1 = p + iq$ and
\begin{align}
	L_- &= -\frac{1}{2}\frac{d^2}{dx^2} + V(x) + g(x)|u_{MAW}|^2 - \mu \,,
\notag \\
	L_+ &= -\frac{1}{2}\frac{d^2}{dx^2} + V(x) + 3g(x)|u_{MAW}|^2 - \mu \,.
\label{bog}
\end{align}
Recall that in the absence of a linear optical lattice, $V(x) \equiv 0$.

In Ref.~\cite{zoilong}, it is shown that a sufficient condition for the onset of the modulational instability is for $\lambda = 0$ to be in the interior of a band of the $L_+$ operator. When $\lambda = 0$, Eqs.~(\ref%
{bog}) decouple, so that $L_+p = L_-q = 0$. With the sinusoidal solution (%
\ref{rr}), we obtain the following Mathieu equations for $V(x) = 0$:
\begin{align}
	-\frac{1}{2}\frac{d^2q}{dx^2} + \left[\left(\frac{\varepsilon c_*^2}{4} -
\mu\right) + \frac{\varepsilon c_*^2}{4}\cos(2\kappa x + 2\varphi_*)\right]
q &= 0\,,  \notag \\
	-\frac{1}{2}\frac{d^2p}{dx^2} + \left[\left(\frac{3\varepsilon c_*^2}{4} -
\mu\right) + \frac{3\varepsilon c_*^2}{4}\cos(2\kappa x + 2\varphi_*)\right]
p &= 0\,,  \label{math}
\end{align}
where we recall that $2\tilde{\mu}=2\mu +(3/8)\varepsilon ^{2} = \kappa^2$, $%
c_{\ast }=\sqrt{ 8\delta _{1}\kappa /3}$, and $\delta_1$ and $\varphi_*$ are
free parameters. (Note that the factors of $g(x)$ in (\ref{bog}) cancel out
with $g(x)$ factors in $|u_{MAW}|^2$ because of the transformation used to
construct the MAW solutions.) In the Mathieu equation $L_+ p = 0$, the
parameter combinations are $3\varepsilon c_*^2/4 = 2\varepsilon\delta_1\kappa $ and $\mu = \kappa^2/2 - 3\varepsilon^2/16$. In
the limit $\varepsilon \longrightarrow 0$ (i.e., $V_0 \longrightarrow 0$),
equations (\ref{math}) become linear harmonic oscillators.

\begin{figure}[tbp]
\centerline{ \includegraphics[width =
0.4\textwidth]{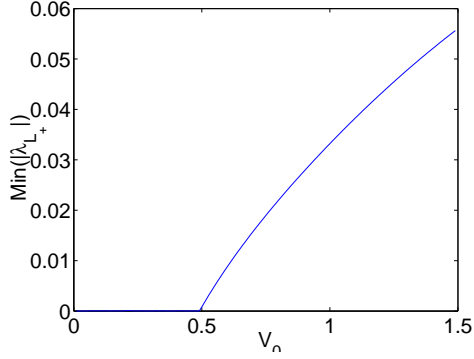}
}
\caption{The minimal-magnitude eigenvalue of $L_+$ as a function of $V_0$
for $\protect\phi_{\star}=0$ solutions. The $0$ value of this diagnostic
indicates the presence of modulational instability.}
\label{full2a}
\end{figure}

To examine the applicability of the aforementioned theorem of \cite{zoilong}
in the present setting, we have computed the eigenvalues of $L_{+}$. In
particular, as a systematic measure for the inclusion of the zero eigenvalue
in a band of eigenvalues of $L_{+}$, we consider $\min \left\{ |\lambda
_{L_{+}}|\right\} $, where $\lambda _{L_{+}}$ denotes the eigenvalues of the
operator. We show this diagnostic in Fig.~\ref{full2a} for solutions with $%
\phi _{\star }=0$.
One can see that for $V_{0}<0.5$ (roughly), the $0$ eigenvalue is, within
the levels of our numerical accuracy, contained in a band of $L_{+}$. This,
according to the results of \cite{zoilong}, implies the existence of
modulational instability in this case. On the other hand, for larger $V_{0}$%
, this no longer occurs.  However, because the criterion only offers a
sufficient condition, the results are inconclusive in this latter case.
Nevertheless, the full numerical results of Fig. \ref{full1} indicate the
presence of instability here as well. For $\phi _{\star }=\pi /2$, we do
\textit{not} find the $0$ eigenvalue remaining in a band of the $L_{+}$
eigenvalues for increasing $V_{0}$.  Hence, for that case as well, we need to
revert to the full numerical results of Fig.~\ref{full2}.

One can also examine the sufficient condition of \cite{zoilong} from a
theoretical perspective using properties of Hill's equations. The eigenvalue
problem $L_+p = \lambda_{L_+} p$ is described by the Mathieu equation
\begin{equation}
	\frac{d^2p}{d\chi^2} + \left[\alpha + 2\beta\cos(2\chi + 2\phi_*)\right] p =
	0\,,  \label{matmat}
\end{equation}
where $\chi = \kappa x$ and
\begin{equation}
	\alpha = 1 + \frac{2\lambda_{L_+}}{\kappa^2} - \frac{4\varepsilon \delta_1}{%
	\kappa} - \frac{3\varepsilon^2}{8 \kappa^2} \,, \quad \beta = -\frac{%
	2\varepsilon\delta_1}{\kappa}\,.
\end{equation}
From Floquet theory, we construct solutions to (\ref{matmat}) of the form $%
p(\chi) = \exp(\gamma(\lambda_{L_+})\chi)Z(\chi)$, where $Z(\chi)$ is a
periodic function with period $\pi$ and $\gamma(\lambda_{L_+})$ is known as
the characteristic (or Floquet) exponent \cite{ince,nayfeh}.

Expanding $Z(\chi)$ in a Fourier series and equating each of its
coefficients to zero yields an infinite-dimensional matrix, whose
determinant $\Delta(\gamma)$ (the so-called ``Hill determinant") is given by
\begin{equation}
	\Delta(\gamma) = \Delta(0) - \frac{\sin^2\left[\frac{1}{2}i\pi\gamma\right]}{%
	\sin^2\left[\frac{1}{2}\pi\sqrt{\alpha}\right]}\,.
\end{equation}
The Floquet exponents satisfy $\Delta(\gamma) = 0$, so they are given by
\cite{ince,nayfeh}
\begin{equation}
	\gamma = \pm \frac{2i}{\pi}\sin^{-1}\left\{\left[\Delta(0)\sin^2\left(\frac{1%
	}{2}\pi\sqrt{\alpha}\right)\right]^{1/2}\right\}\,.
\end{equation}
To determine the spectral bands, one then computes the trace of the
fundamental matrix of (\ref{matmat}), which is given by \cite{pel}
\begin{equation}
	\mathrm{tr}_M(\lambda_{L_+}) = 2\cos\left[-i\pi\gamma(\lambda_{L_+})\right]%
\,.
\end{equation}
The spectral bands of (\ref{matmat}) are defined by the condition $|\mathrm{%
tr}_M(\lambda_{L_+})| \leq 2$.

\begin{figure}[tbp]
\centerline{ \includegraphics[width =0.4\textwidth]{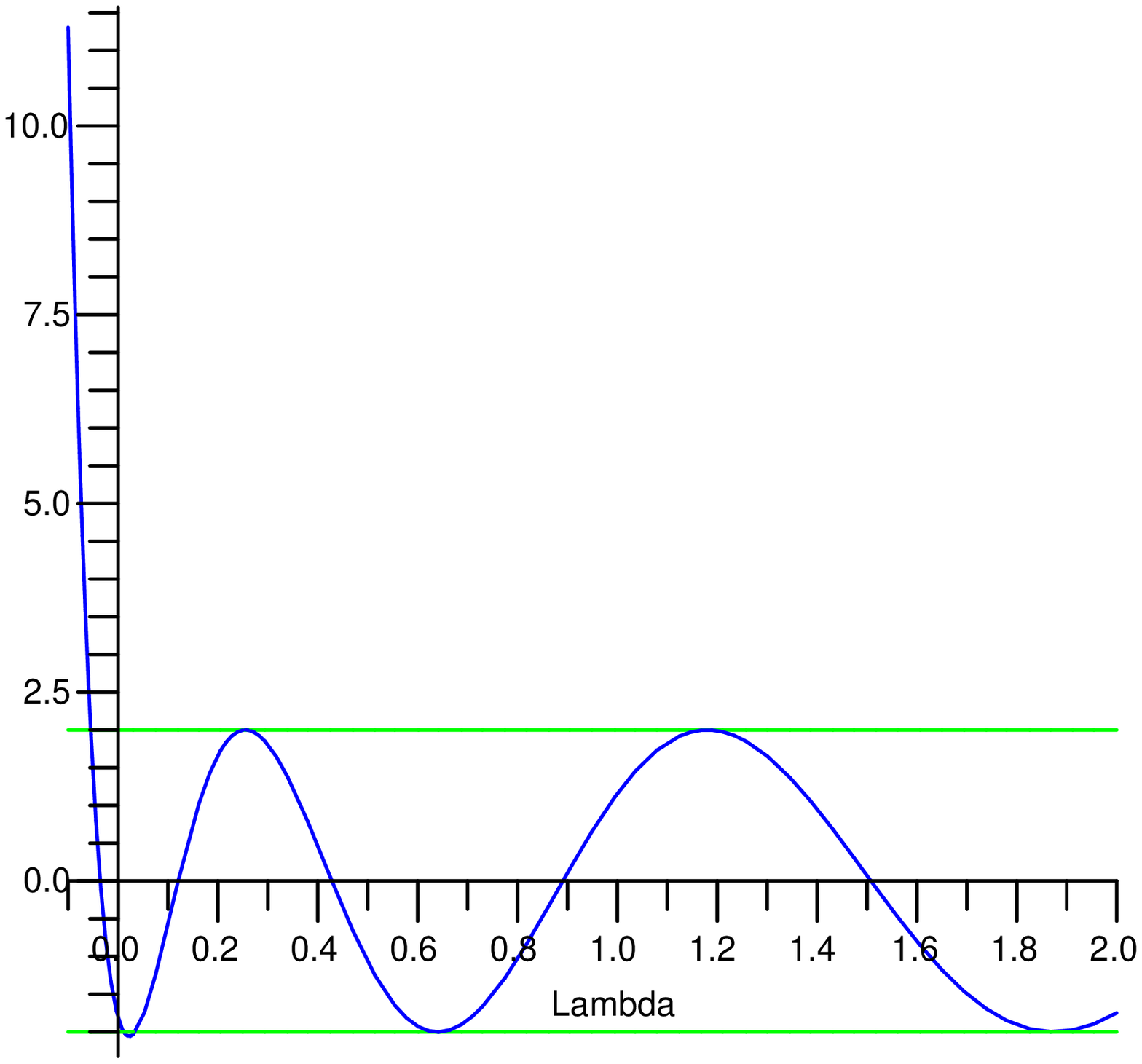}
 \includegraphics[width =0.4\textwidth]{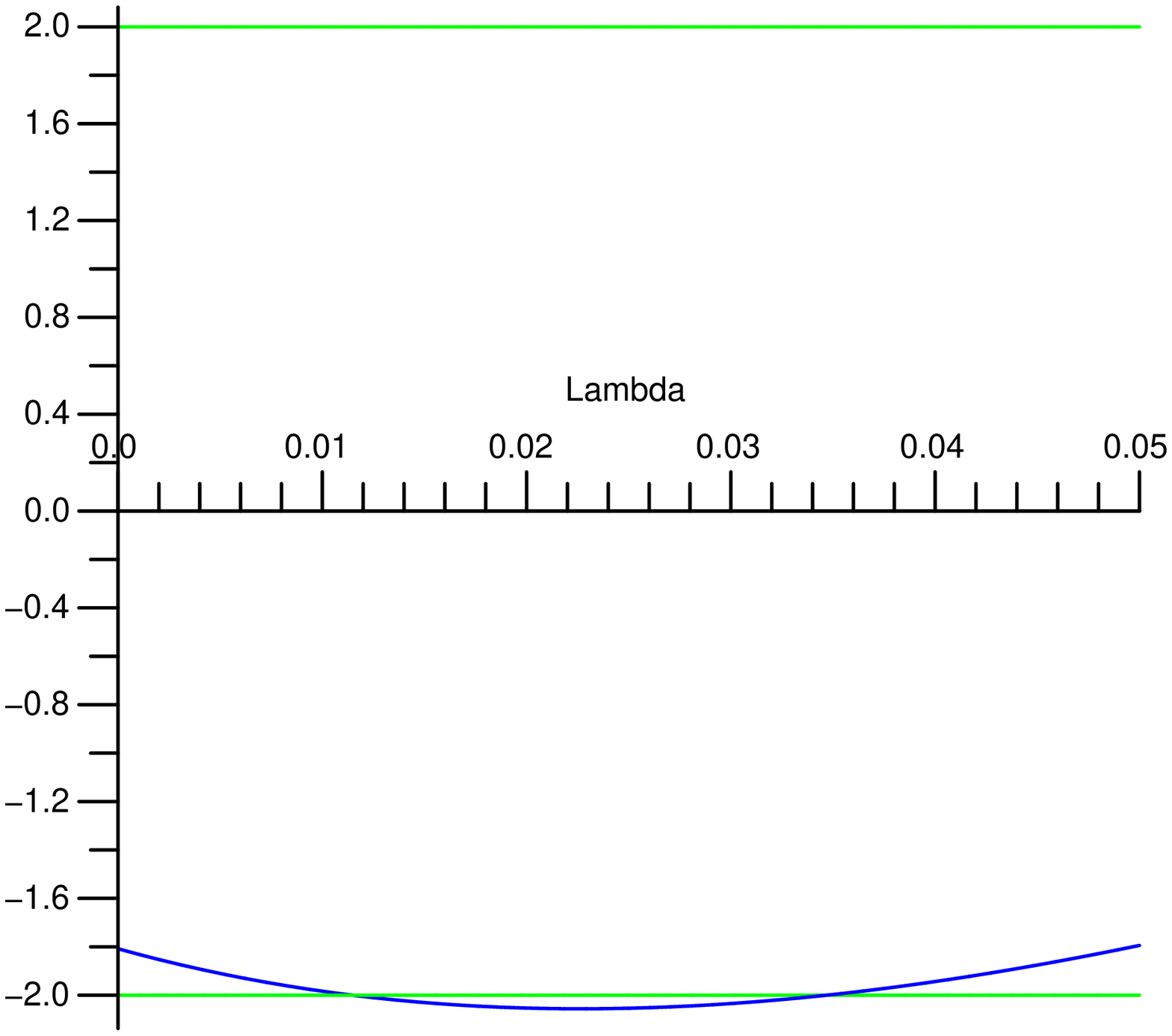}
}
\caption{(Color online) Trace of the fundamental matrix (vertical axis)
versus eigenvalue $\protect\lambda_{L_+}$ of the $L_+$ operator. One obtains
spectral bands when the magnitude of the trace is bounded by 2. The zero
eigenvalue occurs within such a band, indicating that there is a
modulational instability. The right panel shows a magnification of the left
panel.}
\label{hillpic}
\end{figure}

Computing $\Delta(0)$ is, in general, nontrivial, but it is permissible to
consider a finite-dimensional truncation of the (center of the) Hill matrix
provided $\beta$ is small. Using the example $\varphi_* = 0$, the
five-dimensional truncation of the matrix is given by
\begin{equation*}
	\begin{pmatrix}
1 & \frac{\beta}{\alpha - 16} & 0 & 0 & 0 \\
\frac{\beta}{\alpha - 4} & 1 & \frac{\beta}{\alpha - 4} & 0 & 0 \\
0 & \frac{\beta}{\alpha} & 1 & \frac{\beta}{\alpha} & 0 \\
0 & 0 & \frac{\beta}{\alpha - 4} & 1 & \frac{\beta}{\alpha - 4} \\
0 & 0 & 0 & \frac{\beta}{\alpha - 16} & 1%
	\end{pmatrix}%
\,.
\end{equation*}
With the parameter values $\kappa = \pi/8$, $V_0 = 0.15$, $g = 2$, and $%
\delta_1 = 1$ (for which $\beta = 0.15$), we obtain the plot of $\mathrm{tr}%
_M(\lambda_{L_+})$ shown in Fig.~\ref{hillpic}. The eigenvalue $%
\lambda_{L_+} = 0$ is part of the band, indicating that a modulational
instability occurs. While this semi-analytical approach is presented for
completeness and is of interest in its own right, its use of a truncated
Hill matrix and of an approximate analytical solution seems to indicate that
it is preferable to follow the diagnostic of Fig. \ref{full2a} presented
above.

\section{Conclusions}

In this paper, we have investigated the existence and stability of modulated
amplitude waves (MAWs) in quasi-one-dimensional Bose-Einstein condensates (BECs) in
``nonlinear lattices". In particular, we considered an
experimentally feasible situation in which the condensate's
$s$-wave scattering length is modulated periodically in space. Accordingly, we
analyzed the Gross-Pitaevskii (GP) equation with a
spatially periodic nonlinearity coefficient.  We transformed this GP equation into a new GP equation with a constant nonlinearity coefficient and (for small modulations of the scattering length)
an effective (quasi-)superlattice potential. We subsequently studied spatially
extended solutions by applying a coherent-structure ansatz to the
latter equation, which leads to a nonlinear generalized Ince
equation governing the amplitude's spatial dynamics. In the
small-amplitude limit, we used averaging to construct MAW solutions, whose stability we examined using both direct numerical simulations of the original GP equation and
fixed-point computations with the MAWs as numerically
exact solutions.

While direct simulations suggest stability of the constructed
MAW solutions for sufficiently weak periodic modulations, the
fixed-point computations indicate that even those are weakly
unstable (though they can persist long enough to permit
experimental observations) for ``off-site" solutions, whose maxima
do not coincide with maxima of the nonlinear lattice. On the other
hand, ``on-site" MAW solutions are stable for much longer times
even for large periodic modulations of the scattering length. 
This may render the latter solutions more straightforward to
excite and sustain/observe in an experiment in comparison to the
former (even though both exist for wide regimes of the relevant
parameters).  The findings are consistent with recent theoretical work in
\cite{zoilong} on NLS equations with a periodic potential and
constant nonlinearity coefficient, suggesting that the results
reported in that work can be extended to models with spatial
modulation of the nonlinearity, such as the one studied in the
present work.

\textbf{Acknowledgements:} We thank Jared Bronski and Richard Rand for
useful comments during the preparation of this paper. M.A.P. acknowledges
support from the Gordon and Betty Moore Foundation through Caltech's Center
for the Physics of Information. P.G.K. acknowledges funding from
NSF-DMS-0204585, NSF-CAREER, and NSF-DMS-0505663. B.A.M. appreciates partial support from the Israel Science Foundation through the
Excellence-Center grant No. 8006/03. D.J.F. acknowledges partial support
from the Special Research Account of the University of Athens.

\section{Appendix: Functions in the Second-Order Averaging} \label{functions}

In this Appendix, we display some of the functions used in the averaging method in
the main text. The generating functions at first order for the resonant case are
\begin{eqnarray}
	w_{1}(c,\varphi ,x)&=& \frac{c}{2\kappa }\left( \frac{c^{2}}{4\kappa }%
-\delta _{1}\right) \cos (2\kappa x+2\varphi )+\frac{c}{8\kappa }\cos
(4\kappa x+2\varphi )  \notag \\
	&+&\frac{c^{3}}{32\kappa ^{2}}\cos (4\kappa x+4\varphi ) -\frac{c}{ 4\kappa }%
\cos (2\kappa x)\,,  \label{aeq1}
\end{eqnarray}
\begin{eqnarray}
	w_{2}(c,\varphi ,x)& =&\frac{1}{2\kappa }\left( \delta _{1}-\frac{c^{2}}{
2\kappa }\right) \sin (2\kappa x+2\varphi )-\frac{1}{8\kappa }\sin (4\kappa
x+2\varphi)  \notag \\
	&-&\frac{c^{2}}{32\kappa ^{2}}\sin (4\kappa x+4\varphi )-\frac{1}{ 4\kappa }%
\sin (2\kappa x)\,.  \label{aeq2}
\end{eqnarray}
The functions $L_{1}$ and $L_{2}$ that appear at second order are given by
\begin{eqnarray}
	L_{1}(c,\varphi ,x)=&-&\frac{1}{\kappa }\sin (\kappa x+\varphi )\cos (\kappa
x+\varphi )w_{1}D_{1}G_{1}+\frac{c}{\kappa }\sin ^{2}(\kappa x+\varphi
)w_{2}D_{1}G_{1}  \notag \\
	&+&\sin ^{2}(\kappa x+\varphi )w_{1}D_{2}G_{1} +c\sin (\kappa x+\varphi
)\cos (\kappa x+\varphi )w_{2}D_{2}G_{1}  \notag \\
&-& \frac{1}{\kappa }\cos (\kappa x+\varphi )w_{2}G_{1}-\frac{1}{\kappa }%
\sin (\kappa x+\varphi )G_{2}  \label{aeq3}
\end{eqnarray}
\begin{eqnarray}
	L_{2}(c,\varphi ,x) =&-&\frac{1}{\kappa c}\cos ^{2}(\kappa x+\varphi
)w_{1}D_{1}G_{1}+\frac{1}{\kappa }\cos (\kappa x+\varphi )\sin (\kappa
x+\varphi )w_{2}D_{1}G_{1}  \notag \\
	&+&\frac{1}{c}\cos (\kappa x+\varphi )\sin (\kappa x+\varphi
)w_{1}D_{2}G_{1} +\cos ^{2}(\kappa x+\varphi )w_{2}D_{2}G_{1}  \notag \\
&+&\frac{1}{\kappa c}\sin (\kappa x+\varphi )w_{2}G_{1}+\frac{1}{\kappa c^{2}%
}\cos (\kappa x+\varphi )w_{1}G_{1}-\frac{1}{\kappa c}\cos (\kappa c+\varphi
)G_{2}\,,  \label{aeq4}
\end{eqnarray}
where $D_{1}G_{1}$ and $D_{2}G_{1}$ denote, respectively, the derivatives of
$G_{1}(s,s^{\prime },x)$ with respect to $s$ and $s^{\prime }$.

The functions $v_{1}$ and $v_{2}$ are obtained by integrating the terms in $%
L_{1}$ and $L_{2}$ that can be averaged out [i.e., the ones that do not
appear in Eq. (\ref{second})]. Note that $L_{2}$ has seven terms, whereas $%
L_{1}$ has only six. The penultimate term in the former is due to the
presence of $\rho$ in the denominator of the expression for $\phi ^{\prime }$
and arises at second order in the Taylor series because $\rho =c+\varepsilon
w_{1}$.


\begin{thebibliography}{99}
\bibitem{sulem} C. Sulem and P.-L. Sulem, \textit{The Nonlinear Schr\"{o}%
dinger Equation: Self-Focusing and Wave Collapse} (Springer-Verlag, New
York, NY, 1999).

\bibitem{book2} L. Pitaevskii and S. Stringari, \textit{Bose-Einstein
Condensation} (Clarendon Press, Oxford, 2003).

\bibitem{book1} Yu. S. Kivshar and G. P. Agrawal, \textit{Optical Solitons:
From Fibers to Photonic Crystals} (Academic Press, San Diego, 2003).

\bibitem{Folman} R.\ Folman, P.\ Krueger, J.\ Schmiedmayer, J.\ Denschlag
and C.\ Henkel, Adv.\ Atom.\ Mol.\ Opt.\ Phys.\ \textbf{48}, 263 (2002); J.
Reichel, Appl. Phys. B \textbf{75}, 469 (2002); J. Fortagh and C.
Zimmermann, Science \textbf{307} 860 (2005).

\bibitem{Weiner03} J. \ Weiner, \textit{Cold and Ultracold Collisions in
Quantum Microscopic and Mesoscopic Systems}, Cambridge University Press 2003.

\bibitem{feshbachNa} S. Inouye, M. R. Andrews, J. Stenger, H. J. Miesner, D.
M. Stamper-Kurn and W. Ketterle, Nature \textbf{392}, 151 (1998); J.
Stenger, S. Inouye, M. R. Andrews, H.-J. Miesner, D. M. Stamper-Kurn, and W.
Ketterle, Phys. Rev. Lett. \textbf{82}, 2422 (1999).

\bibitem{feshbachRb} J. L. Roberts, N. R. Claussen, J. P. Burke, Jr., C. H.
Greene, E. A. Cornell, and C. E. Wieman, Phys. Rev. Lett. \textbf{81}, 5109
(1998); S. L. Cornish, N. R. Claussen, J. L. Roberts, E. A. Cornell, and C.
E. Wieman, Phys. Rev. Lett. \textbf{85}, 1795 (2000).

\bibitem{Theis04} M. Theis, G. Thalhammer, K. Winkler, M. Hellwig, G. Ruff,
R. Grimm, and J. H. Denschlag, Phys. Rev. Lett. \textbf{93}, 123001 (2004).

\bibitem{electric} L. You and M. Marinescu, Phys. Rev. Lett. \textbf{81},
4596 (1998).

\bibitem{olshani} M. \ Olshanii, Phys. Rev. Lett. \textbf{81}, 938 (1998);
T. Bergeman, M. G. Moore and M. Olshanii, Phys. Rev. Lett. \textbf{91},
163201 (2003).

\bibitem{Napoli} S. De Nicola, B. A. Malomed, and R. Fedele, Phys. Lett. A
\textbf{360}, 164 (2006).

\bibitem{molecule} J. Herbig, T. Kraemer, M. Mark, T. Weber, C. Chin, H. C.
Nagerl, and R. Grimm, Science \textbf{301}, 1510 (2003); C. A. Regal, C.
Ticknor, J. L. Bohn, and D. S. Jin, Nature \textbf{424}, 47 (2003).

\bibitem{becbcs} M. Bartenstein, A. Altmeyer, S. Riedl, S. Jochim, C. Chin,
J. H. Denschlag, and R. Grimm, Phys. Rev. Lett. \textbf{92}, 203201 (2004).

\bibitem{FRM1} F. Kh. Abdullaev, J. G. Caputo, R. A. Kraenkel, and B. A.
Malomed Phys. Rev. A \textbf{67}, 013605 (2003); H. Saito and M. Ueda, Phys.
Rev. Lett. \textbf{90}, 040403 (2003); G. D. Montesinos, V. M. P\'erez-Garc%
\'{\i}a, and P. J. Torres, Physica D \textbf{191}, 193 (2004).

\bibitem{FRMr} M. Matuszewski, E. Infeld, B. A. Malomed, and M. Trippenbach,
Phys. Rev. Lett. \textbf{95}, 050403 (2005); M. Trippenbach, M. Matuszewski,
and B. A. Malomed, Europhys. Lett. \textbf{70}, 8 (2005).

\bibitem{FRM2} P. G. Kevrekidis, G. Theocharis, D. J. Frantzeskakis, and B.
A. Malomed, Phys. Rev. Lett. \textbf{90}, 230401 (2003); D. E. Pelinovsky,
P. G. Kevrekidis, and D. J. Frantzeskakis, Phys. Rev. Lett. \textbf{91},
240201 (2003); F. Kh. Abdullaev, E. N. Tsoy, B. A. Malomed, and R. A.
Kraenkel, Phys. Rev. A \textbf{68}, 053606 (2003); F. Kh. Abdullaev, A. M.
Kamchatnov, V. V. Konotop, and V. A. Brazhnyi Phys. Rev. Lett. \textbf{90},
230402 (2003); Z. Rapti, G. Theocharis, P. G. Kevrekidis, D. J.
Frantzeskakis and B. A. Malomed, Phys. Scripta \textbf{T107}, 27 (2004); D.
E. Pelinovsky, P. G. Kevrekidis, D. J. Frantzeskakis, and V. Zharnitsky,
Phys. Rev. E \textbf{70}, 047604 (2004); Z. X. Liang, Z. D. Zhang, and W. M.
Liu, Phys. Rev. Lett. \textbf{94}, 050402 (2005). A. Gubeskys, B. A.
Malomed, and I. M. Merhasin, Stud. Appl. Math. \textbf{115}, 255 (2005); M.
A. Porter, M. Chugunova, and D. E. Pelinovsky, Phys. Rev. E {\bf 74}, 036610 (2006). 

\bibitem{Kestutis} K. Staliunas, S. Longhi, and G. J. de Valcarcel, Phys.
Rev. Lett. \textbf{89}, 210406 (2002).

\bibitem{fka} F. Kh. Abdullaev and M. Salerno, J. Phys. B \textbf{36}, 2851
(2003).

\bibitem{fermi} H. Xiong, S. Liu, M. Zhan, and W. Zhang, Phys. Rev. Lett.
\textbf{95}, 120401 (2005).

\bibitem{g1} G. Theocharis, P. Schmelcher, P. G. Kevrekidis, and D. J.
Frantzeskakis, Phys. Rev. A \textbf{72}, 033614 (2005).

\bibitem{v1} M. I. Rodas-Verde, H. Michinel, and V. M. P\'erez-Garc\'{\i}a,
Phys. Rev. Lett. \textbf{95}, 153903 (2005).


\bibitem{Peter} P. Y. P. Chen and B. A. Malomed, J. Phys. B: At. Mol. Opt.
Phys. \textbf{38}, 4221 (2005); \textbf{39}, 2803 (2006).


\bibitem{v2} A. V. Carpentier, H. Michinel, M. I. Rodas-Verde, and V. M. P%
\'{e}rez-Garc\'{\i}a, Phys. Rev. A 74, 013619 (2006).

\bibitem{g2} G. Theocharis, P. Schmelcher, P. G. Kevrekidis, and D. J.
Frantzeskakis, Phys. Rev. A {\bf 74}, 053614 (2006). 

\bibitem{fka2} J. Garnier and F. Kh. Abdullaev, Phys. Rev. A \textbf{74},
013604 (2006).

\bibitem{fka3} F. Kh. Abdullaev, and J. Garnier, Phys. Rev. A \textbf{72},
061605(R) (2005).

\bibitem{malspace} H. Sakaguchi and B. A. Malomed, Phys. Rev. E \textbf{73},
026601 (2006).

\bibitem{Gadi} G. Fibich and X.-P. Wang, Physica D \textbf{175}, 96 (2003).

\bibitem{fka4} F. Kh. Abdullaev, A. Gammal, and L. Tomio, J. Phys. B \textbf{%
37}, 635 (2004).

\bibitem{Manakov}  R. K. Bullough, A. P. Fordy, and S. V. Manakov, Phys.
Lett. A \textbf{91}, 98 (1982).


\bibitem{quasibec} S. Peil, J. V. Porto, B. Laburthe Tolra, J. M. Obrecht,
B. E. King, M. Subbotin, S. L. Rolston, and W. D. Phillips, Phys. Rev. A
\textbf{67}, 051603 (2003).

\bibitem{gp1d} V. M.\ P\'{e}rez-Garc\'{\i}a, H.\ Michinel, and H.\ Herrero,
Phys.\ Rev.\ A \textbf{57}, 3837 (1998); Yu. S. Kivshar, T. J. Alexander,
and S. K. Turitsyn, Phys. Lett. A \textbf{278}, 225 (2001); L.\ Salasnich,
A.\ Parola and L.\ Reatto, Phys.\ Rev.\ A \textbf{65}, 043614 (2002); Y. B.\
Band, I.\ Towers, and B. A.\ Malomed, Phys.\ Rev.\ A \textbf{67}, 023602
(2003).

\bibitem{map1} M. A. Porter and P. Cvitanovi\'c, Phys. Rev. E \textbf{69},
047201 (2004).

\bibitem{map2} M. A. Porter and P. Cvitanovi\'c, Chaos \textbf{14}, 739
(2004).

\bibitem{map3} M. A. Porter and P. G. Kevrekidis, SIAM J. App. Dyn.
Sys. \textbf{4}, 783 (2005).

\bibitem{map4} M. A. Porter, P. G. Kevrekidis, and B.A. Malomed, Physica D
\textbf{196}, 106 (2004).

\bibitem{ngince} L. Ng and R. Rand, Nonlinear Dynamics \textbf{31}, 73
(2003).


\bibitem{675} R. H. Rand, {\it Lecture Notes on Nonlinear Vibrations}, available online at {\it http://www.tam.cornell.edu/randdocs/nlvibe45.pdf}.


\bibitem{ngmat} L. Ng and R. Rand, Chaos, Solitons and Fractals \textbf{14},
173 (2002).

\bibitem{bernard} B. Deconinck and J. N. Kutz,
J. Comp. Physics, \textbf{219} 296 (2006).

\bibitem{wein06} G. Fibich, Y. Sabin, and M. I. Weinstein, Physica D \textbf{%
175}, 96 (2003).

\bibitem{zoilong} J. C. Bronski and Z. Rapti, Dynamics of PDEs \textbf{2},
335 (2005).

\bibitem{ince} E. L. Ince, \textit{Ordinary Differential Equations} (Dover
Publications, Inc., New York, NY, 1956).

\bibitem{nayfeh} A. H. Nayfeh and D. T. Mook, \textit{Nonlinear Oscillations}
(John Wiley \& Sons, Inc., New York, NY, 1995).

\bibitem{pel} D. E. Pelinovsky, A. A. Sukhorukov, and Yu. S. Kivshar, Phys.
Rev. E \textbf{70}, 036618 (2004).
\end{thebibliography}

\end{document}